%% file: sn-article.tex
\documentclass[pdflatex,sn-mathphys,Numbered]{sn-jnl}

\usepackage{url}
\usepackage{graphicx}%
\usepackage{multirow}%
\usepackage{amsmath,amssymb,amsfonts}%
\usepackage{amsthm}%
\usepackage{mathrsfs}%
\usepackage[title]{appendix}%
\usepackage{xcolor}%
\usepackage{textcomp}%
\usepackage{manyfoot}%
\usepackage{booktabs}%
\usepackage{algorithm}%
\usepackage{algorithmicx}%
\usepackage{algpseudocode}%
\usepackage{listings}%
\usepackage{upgreek}%
\usepackage{soul}
\usepackage{siunitx}
\usepackage{lmodern}
\usepackage{ulem}  
\usepackage{subcaption}

\soulregister\cite7
\soulregister\ref7
\begin{document}
\title[Article Title]{A New Method for Identifying Contaminating Sources and Locating Target Sources through the Cross-Arm Features of Micro Pore Optics}


\author[1,2]{\fnm{Yiming} \sur{Huang}}
\author[1]{\fnm{Lian} \sur{Tao}}\email{taolian@ihep.ac.cn}
\author[1]{\fnm{Jin-Yuan} \sur{Liao}}\email{liaojinyuan@ihep.ac.cn}
\author[1,2,3]{\fnm{Shuang-Nan} \sur{Zhang}}\email{zhangsn@ihep.ac.cn}
\author[4]{\fnm{Stéphane} \sur{Schanne}}
\author[4]{\fnm{Bertrand} \sur{Cordier}}
\author[1]{\fnm{Shaolin} \sur{Xiong}}
\author[1]{\fnm{Juan} \sur{Zhang}}
\author[1]{\fnm{Zhengwei} \sur{Li}}
\author[1]{\fnm{Qian-Qing} \sur{Yin}}
\author[1]{\fnm{Xiangyang} \sur{Wen}}
\author[1]{\fnm{Sheng} \sur{Yang}}
\author[1]{\fnm{Min} \sur{Gao}}
\author[3]{\fnm{Donghua} \sur{Zhao}}
\author[1]{\fnm{Xiang} \sur{Ma}}
\author[1]{\fnm{Yue} \sur{Huang}}
\author[1]{\fnm{Liang} \sur{Zhang}}
\author[1]{\fnm{Liming} \sur{Song}}

\affil[1]{\orgdiv{Key Laboratory for Particle Astrophysics, Institute of High Energy Physics}, \orgname{Chinese Academy of Sciences}, \orgaddress{\city{Beijing}, \postcode{100149}, \country{China}}}

\affil[2]{\orgdiv{University of Chinese Academy of Sciences}, \orgname{Chinese Academy of Sciences}, \orgaddress{\city{Beijing}, \postcode{100149}, \country{China}}}

\affil[3]{\orgdiv{National Astronomical Observatories}, \orgname{Chinese Academy of Sciences}, \orgaddress{\city{Beijing}, \postcode{100101}, \country{China}}}

\affil[4]{\orgdiv{IRFU/Departement d’Astrophysique - AIM}, \orgname{CEA Paris Saclay}, \orgaddress{\city{Gif-sur-Yvette}, \postcode{91191}, \country{France}}}

\abstract{The Pathfinder of the Type-A satellites in the Chasing All Transients Constellation Hunters (CATCH) space mission is equipped with Micro-Pore Optics (MPOs) and four single-pixel Silicon Drift Detectors (SDDs). Due to the lack of position resolution in an individual SDD, we propose a new method based on the cross-arms in the point spread function (PSF) of MPOs to enhance the satellite's capability in identifying contaminating sources and locating target sources. By placing one detector on each of the horizontal and vertical cross-arms on the focal plane, we can use the changes in the relative counts on the cross-arms detectors to deduce the location of the source. Simulated observations demonstrate that, for a target source with a flux of 1\,Crab and an exposure time of 200\,s, the cross-arms detectors can identify contaminating source with the same flux level at an off-axis angle larger than $\ang{;8;}$, and improve positioning accuracy to $\ang{;6;}$. Furthermore, we extend the simulation study to CATCH Type-A, which plans to use an SDD array. In situations where sources exhibit the same flux of 1\,Crab and the exposure time is merely 1\,s, a 16$\times$16 SDD array is capable of identifying contaminating source with an off-axis angle greater than \ang{;2.4;} and can achieve a positioning precision of $\ang{;1.8;}$. }


\maketitle
\section{Introduction}\label{sec1}

In the field of time-domain astronomy, a telescope's capabilities in identifying contaminating sources and locating target sources are of vital importance. The improvement in positioning capability enables more accurate and effective triggering, permitting timely multi-wavelength follow-up observations. Identifying contaminating sources helps remove their interference with the target sources, enabling the capture of the targets' intrinsic properties. These capabilities advance our understanding of the underlying astrophysics of various objects.
   
The observing capability of an X-ray astronomical satellite primarily depends on the combination of its focusing mirror and detector. Currently, Chandra boasts the best angular resolution in X-ray astronomy, utilizing a precision-engineered Wolter-\uppercase\expandafter{\romannumeral1} optical system \citep{wolter} and a detector array with small pixel size matching the mirror's angular resolution, achieving an angular resolution of up to $\ang{;;0.5}$ \citep{chandra2000}. However, the long development cycle of Chandra, the high costs associated with the large-area detector array, and the technical challenges of manufacturing such a precise focusing mirror hinder widespread replication of this design. Furthermore, Chandra's focusing mirror weighs nearly 956.4\,kg \citep{chandra}, necessitating the use of a large satellite platform. 
   
Nowadays, numerous transients have been discovered in time-domain astronomy, and the rate of new discoveries is expected to continue to increase rapidly in the future. This trend requires multiple X-ray astronomy satellites to conduct follow-up observations. Given the mass, development cycle, and costs associated with Chandra-like satellites, deploying a significant number to meet the demand for precise observations of numerous transients is impractical. Therefore, adopting small satellite platforms that feature lightweight optics, short development cycles, and lower costs is the inevitable choice for the future. While the angular resolution of these small X-ray telescopes falls short of that of Chandra, they can be deployed in large numbers to meet the needs of follow-up observations of a large number of transients and to identify promising targets for in-depth observations by large observatories.
   
The Chasing All Transients Constellation Hunters (CATCH) space mission was proposed by the Institute of High Energy Physics (IHEP) of the Chinese Academy of Sciences in 2019 \citep{licatch}. It is an intelligent X-ray astronomical satellite constellation mission, which aims to address the severe lack of follow-up observation capabilities in the time-domain astronomy era. By deploying a constellation of over one hundred micro-satellites equipped with compact and lightweight X-ray focusing telescopes, the mission will utilize advanced artificial intelligence-based constellation control technology to achieve a multi-parameter (timing, spectroscopy, imaging, and polarization) characterization of the dynamic universe. 
   
One type of satellites within the CATCH constellation is CATCH Type-A \citep{huang,huangground}. Its scientific objective is to conduct timing monitoring of targets after discovery, incorporating imaging and positioning capabilities. Therefore, it is equipped with a combination of Micro Pore Optics (MPOs) and Silicon Drift Detector (SDD) array. As the multi-pixel SDD array is still under development, the Pathfinder of CATCH Type-A, intended for technical validation, will utilize four single-pixel SDDs. Through simulations, we explore its capabilities in identifying contaminating sources and locating target sources, and proposing an optimized detector layout design. Additionally, the extent to which the use of the SDD array can enhance observing capability is evaluated. This work is not only crucial for the CATCH Type-A Pathfinder, but also has broader implications for identifying contaminating sources and enhancing localization capabilities in other X-ray telescopes that lack imaging capabilities.
   
The paper is structured as follows: Section~\ref{sec2} introduces the configuration of the Pathfinder of CATCH Type-A; Section~\ref{sec3} evaluates the presence of contaminating sources within the field of view (FOV) during observations; Section~\ref{sec4} discusses the ability of the Pathfinder of CATCH Type-A to identify contaminating sources under two detector layouts; Section~\ref{sec5} explores the positioning capability using cross-arms detectors; Section~\ref{sec6} describes the SDD array planned for future and its capabilities to identify contaminating sources and locate target sources; and Section~\ref{sec7} provides a summary.

\section{Pathfinder of CATCH Type-A}\label{sec2}

\begin{figure}[ht]%
\centering
\includegraphics[width=0.78\textwidth]{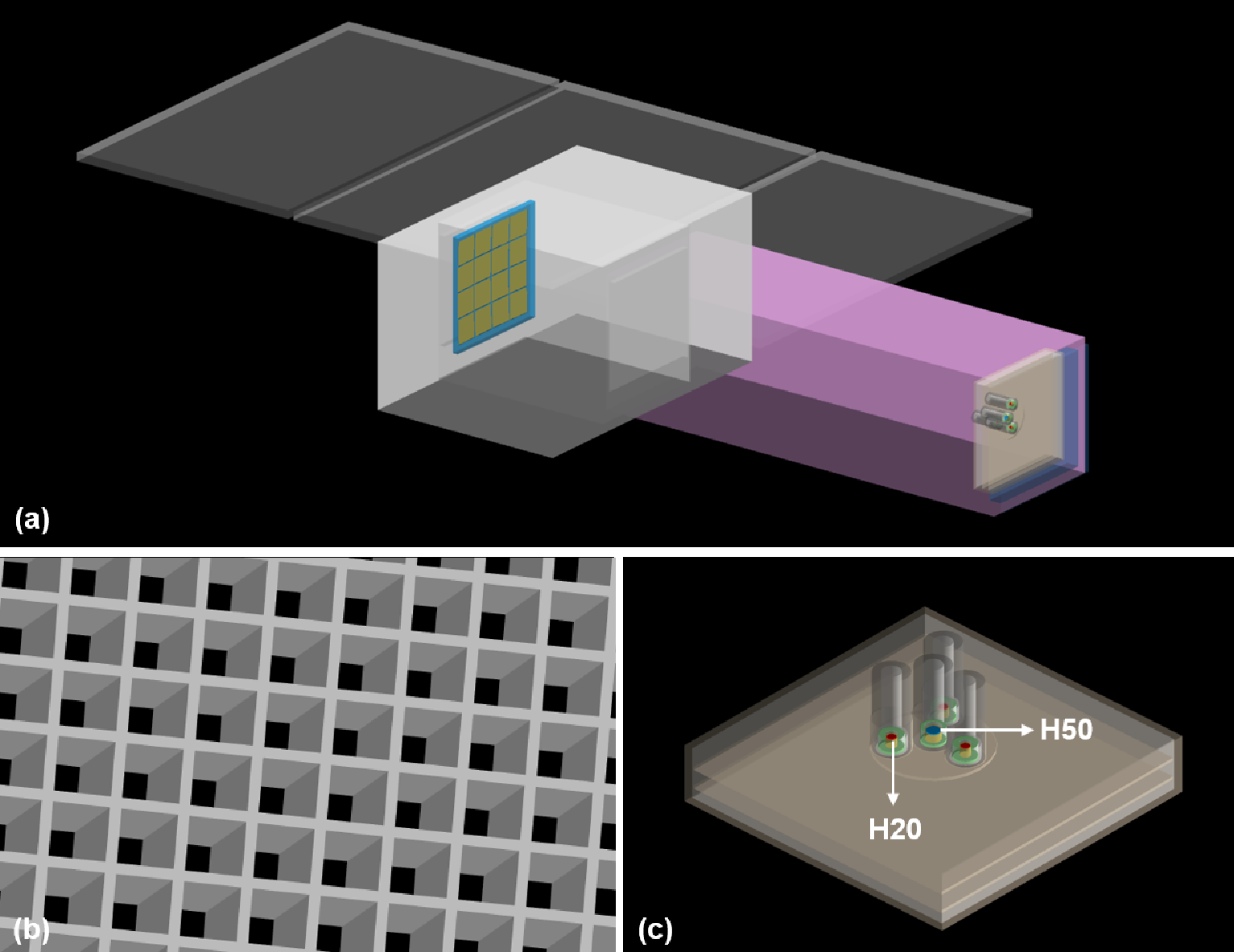}
\caption{Panel (a): Configuration of the Pathfinder of CATCH Type-A. Panel (b): Partial enlarged view of the MPOs. Panel (c): Enlarged view of the detector system.}\label{fig_model}
\end{figure}

The Pathfinder of CATCH Type-A is made up of a platform, a deployable mast, a focusing mirror system, and a detector system. The focusing mirror system is placed in the front of the platform, while the detector system is placed at the end of the deployable mast, 1\,m away from the focusing mirror system. Panel (a) in Figure \ref{fig_model} shows the satellite model constructed in Geant4 \citep{1geant4,2geant4,3geant4}.

\begin{figure}[h]%
\centering
\includegraphics[width=0.55\textwidth]{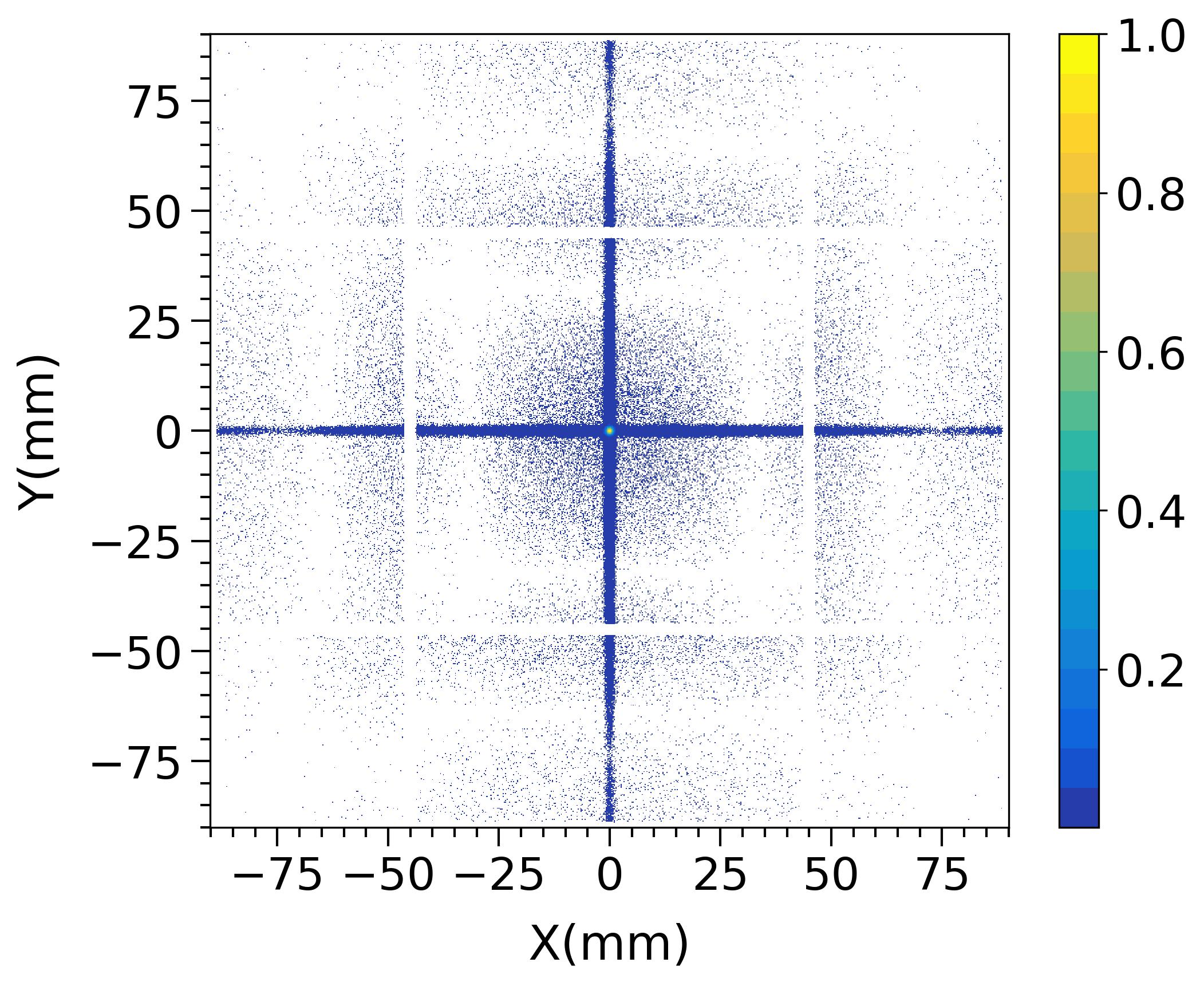}
\caption{Simulated PSF of the MPOs used in the Pathfinder of CATCH Type-A. It comprises a focused spot, horizontal and vertical cross-arms, and some diffuse patches. These arise from different photon reflection numbers: two reflections form the focal spot, an odd number creates the arms, and zero or an even number yields the diffuse patches. The color bar on the right represents the normalized counts. The gaps at $\pm$45\,mm are caused by the shadow of the supporting frame.}\label{fig_psf}
\end{figure}

The focusing mirror is a 4$\times$4 array of light-weight MPOs, with a total mass of approximately 500\,g and an overall size of 200\,mm$\times$200\,mm \citep{xiaoground}. It collects the photon with energy between 0.5\,keV and 4.0\,keV. Each MPO is a spherical shell with a radius of curvature of 2\,m, containing millions of micro square pores that radially point towards a common center (see Panel (b) in Figure \ref{fig_model}). Grazing incidence X-rays are reflected off the walls of these pores and focused onto a spherical focal surface \citep{angel}. This focal surface shares the same center as the MPOs and its curvature is half that of the MPO assembly, i.e., 1\,m. The point spread function (PSF) of this focusing mirror is shown in Figure \ref{fig_psf}. The cross feature formed by MPOs comprises a focal spot, horizontal and vertical cross-arms, and some diffuse patches. This pattern arises because X-ray photons undergoing different numbers of reflections on the micro-pore sidewalls are focused to different locations. Photons that undergo two reflections converge at the focal spot; those experiencing single or successive odd reflections are deflected to the horizontal or vertical arms of the cross; and photons that either pass straight through the pores or undergo multiple even reflections fall into the diffuse regions \cite{lobster}. 

The detector system used in the Pathfinder of CATCH Type-A consists of four SDDs: one H50 detector in the center with a geometric area of 50\,mm$^2$ and three H20 detectors around it with a geometric area of 20\,mm$^2$ each (see Panel (c) in Figure \ref{fig_model}) \citep{lidetector, SDD}. These four detectors are installed on a common planar surface. The central H50 detector is positioned exactly on the ideal spherical focal surface with a radius of curvature of 1\,m, while the surrounding H20 detectors are located up to 23\,mm away from the H50 within this plane. This planar arrangement results in a maximum geometrical defocus of 0.26\,mm relative to the ideal spherical focal surface. The deviation falls well within the mechanical and optical tolerances of this MPO focusing mirror, and its impact is negligible. The sensitive layers of these detectors are 450\,$\upmu$m-thick Silicon, and each detector is surrounded by an Aluminum collimator to limit the background level. The SDD is chosen for its advantages of fast time readout and excellent energy resolution; however, the single SDD pixel lacks position resolution capability and is limited to providing the count. The combination of the central detector and the focusing mirror covers a FOV of $\ang{0.4;;}\times\ang{0.4;;}$. 

The platform adopts a three-axis stabilized attitude control system. During observations, the Sun-avoidance constraint requires that the angle between the sun and the optical axis (defined by the normal direction of the MPOs plane) remains within \ang{25;;}$\sim$\ang{180;;}, ensuring safe operation of the payload. According to annual sky-visibility statistics, any position on the celestial sphere can accumulate more than 100 days of observable time per year. After receiving an observation instruction from the ground station, the satellite can slew toward the target with a maneuver capability of approximately \ang{45;;} in 2 minutes, allowing it to reorient to the target within a few minutes. The pointing accuracy of the platform is \ang{;1;}, with the pointing stability better than $\ang{0.01;;}/\mathrm{s}$. The boresight is aligned with the central H50 detector, so when the satellite points at a target source, the focused spot is expected to fall onto the central detector. Nevertheless, within the $\ang{0.4;;}\times\ang{0.4;;}$ FOV, additional sources may enter the field and act as contaminating sources during observations.

At the constellation level, the current design adopts a Walker constellation configuration with three orbital planes at an orbital altitude of 550\,km and an inclination of \ang{29;;}. Each orbital plane hosts satellites with complementary observational capabilities, enabling coordinated timing, spectroscopy, imaging, and polarization measurements. The observing constraints include an Earth elevation angle larger than \ang{30;;}, a Sun avoidance angle greater than \ang{25;;}, a moon avoidance angle exceeding \ang{15;;}, and a geomagnetic cutoff rigidity larger than 6 in high particle background regions \citep{licatch}. Within these orbital and operational constraints, the constellation design allows multiple satellites to perform relay observations of single or multiple targets, achieving continuous temporal coverage, while also flexibly forming collaborative FOV configurations to monitor key sources. The constellation configuration may be further refined and optimized in future phases.

\section{Analysis of potential contaminating sources}\label{sec3}

\begin{figure}[b]%
\centering
\includegraphics[width=\textwidth]{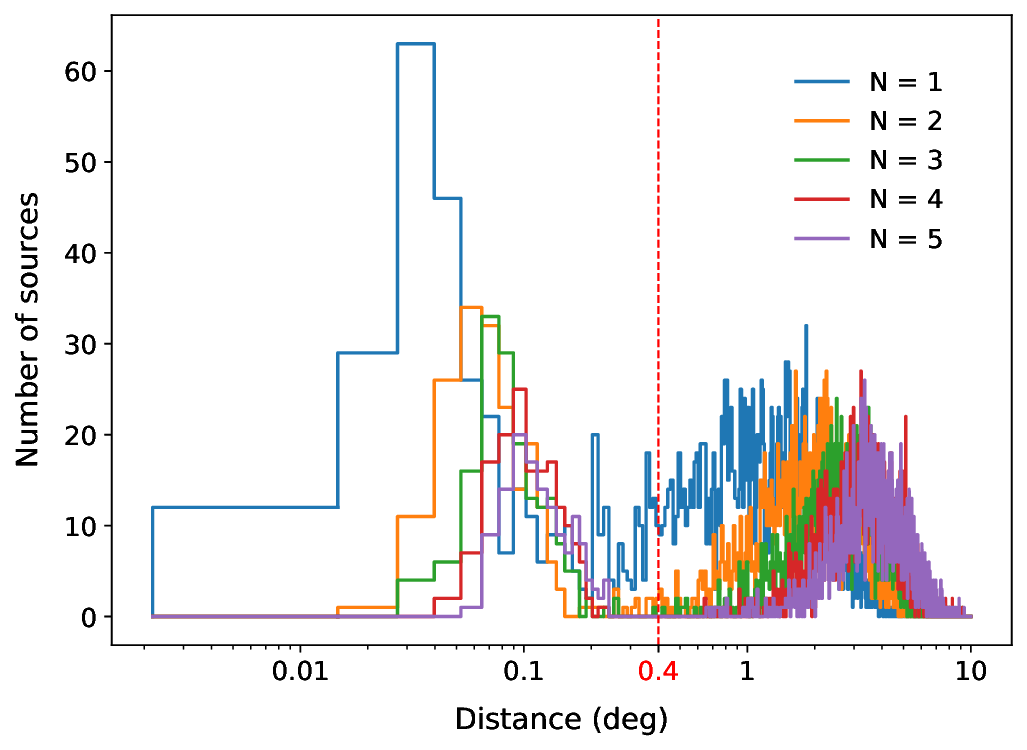}
\caption{Distribution of angular distances to the $N$-th nearest neighbors ($N=1$--$5$). Different colors indicate different neighbor orders, and the red dashed line denotes the FOV. }
\label{fig_Xraycatalog105}
\end{figure}

To evaluate the level of contaminating sources within the FOV, we analyzed the distribution of distances between sources in the Second ROSAT All-Sky Survey Point Source Catalog (2RXS). With a flux limit of approximately $1.0\times10^{-13}$\,erg\,cm$^{-2}$\,s$^{-1}$, this catalog comprises around 135,000 X-ray detections with a detection likelihood above 6.5 in the 0.1--2.4\,keV energy band \citep{2rxs}. The detection likelihood is the statistic generated by the maximum likelihood detection algorithm used in the ROSAT data processing, which quantifies the probability that a detection is real rather than a background fluctuation——higher values indicate more reliable detections. We used the 2RXS catalog because its energy band matches that of the CATCH Type-A Pathfinder (0.5--4 keV).

Based on the CATCH Type-A Pathfinder's sensitivity of $2.4\times10^{-12}$\,erg\,cm$^{-2}$\,s$^{-1}$, we selected 2RXS sources exceeding this flux level, resulting in a total of 3,591 sources. Using the source coordinates (right ascension and declination) from the 2RXS, we computed the angular distances to the first five nearest neighbours for each source, with the statistical results shown in Figure~\ref{fig_Xraycatalog105}. The fractions of sources whose 1st to 5th nearest neighbors fall within the $\ang{0.4;;}\times\ang{0.4;;}$ FOV are 11.7\%, 5.51\%, 4.71\%, 4.46\%, and 3.76\%, respectively. These results indicate that 6.19\% of sources have exactly one contaminating neighbor within the FOV, while 0.80\%, 0.25\%, and 0.70\% contain two, three, and four contaminating neighbors, respectively. Therefore, most contamination events are expected to be caused by a single nearby source, although multi-source contamination remains possible. Given that the single-contaminant scenario is the most common and representative case, this work focuses on the situation where one target source and one contaminating source are present within the FOV. Extensions to multi-source contamination will be explored in future studies.

It is important to note that the 2RXS primarily consists of persistent sources, while transient sources, which are the main targets of CATCH, may involve a more complex situation. CATCH targets various transient events, including X-ray binaries, stellar flares, gamma-ray bursts, fast radio bursts, etc. In practical observations, these transient sources may serve as the intended targets or may enter the FOV as a contaminating source when another target is being observed. And the contaminating source may be a known catalogued object, a newly emerging transient, or a background source. The statistics derived from 2RXS suggest that multiple-source scenarios occur frequently enough across the X-ray sky within the FOV of CATCH Type-A Pathfinder, highlighting the importance of identifying contaminating sources during real observations.


\section{Contaminating source identification}\label{sec4}

To study and improve the performance of the CATCH Type-A Pathfinder in identifying contaminating sources, simulated observations were made using the Monte Carlo simulation software Geant4 \citep{qi, Xraytracing, zhao}. The simulation is based on a complete satellite model including highly detailed MPOs and detector system. The MPOs were modeled with realistic fabrication imperfections, such as pore pointing deviations and coating surface roughness, enabling the simulated PSF to approximate the expected in-flight performance. In this section, we present the initial and updated layouts of the detector system, evaluate the satellite's contaminating source identification performance under each layout, and include a detailed discussion.

\subsection{Initial layout}\label{subsec1}

\begin{figure}[h]%
\centering
\includegraphics[width=0.55\textwidth]{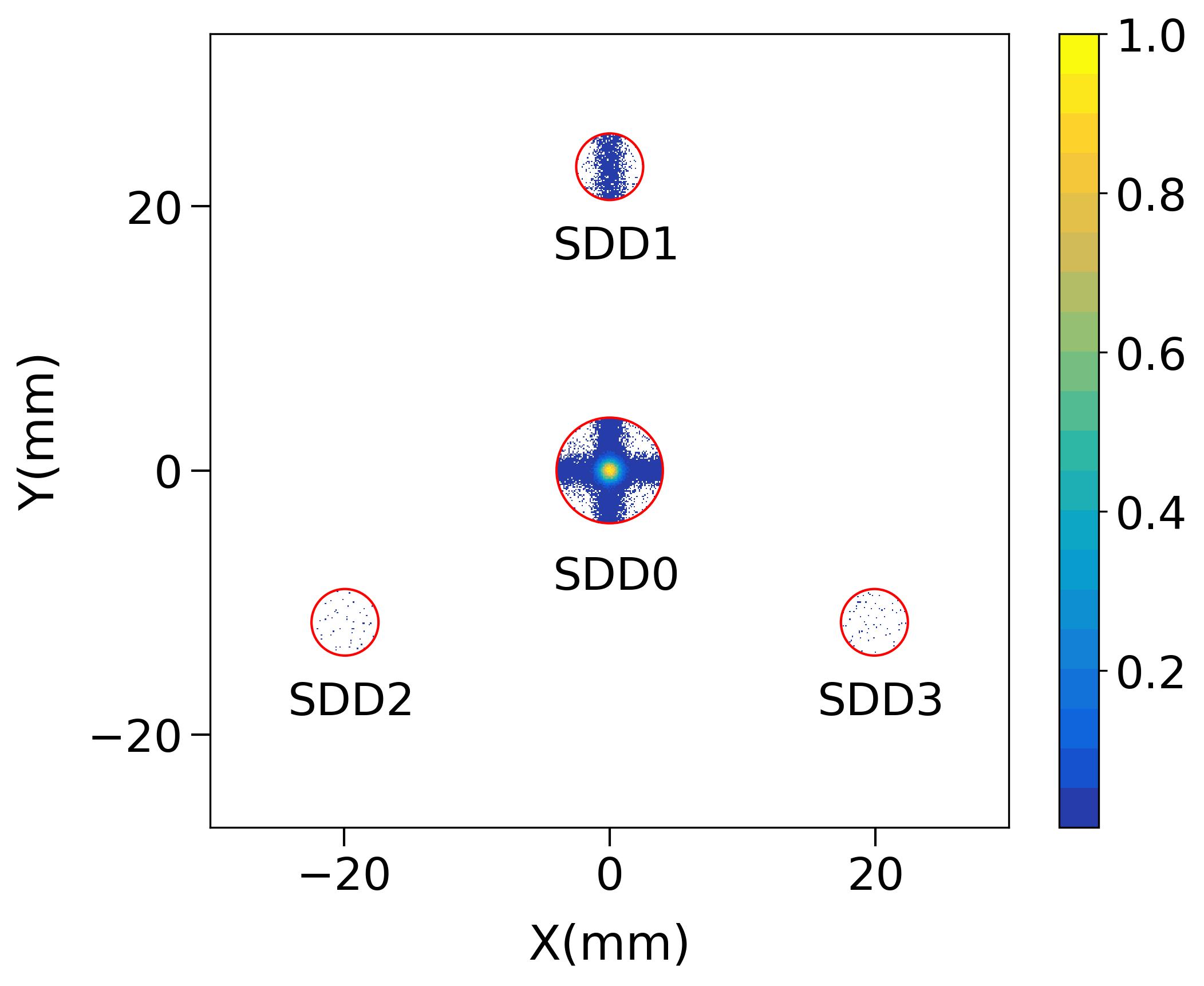}
\caption{Projection of the source PSF onto the detector system plane in the initial layout.}\label{fig_SDDv1psf}
\end{figure}

In the initial design of the detector layout, the scheme is as follows: the H50 detector serves as the central primary detector, labeled as SDD0. Three H20 detectors are evenly distributed around it at mutual angles of \ang{120;;}, functioning as background detectors and labeled as SDD1 (top), SDD2 (left), and SDD3 (right), respectively. Figure \ref{fig_SDDv1psf} illustrates the source PSF projected onto the detector system plane. The figure shows that SDD0 covers the focal spot and partial cross-arms of the PSF. SDD1 covers part of the vertical arm, while SDD2 and SDD3 cover partial diffuse patches.

\begin{figure}[h]%
\centering
\includegraphics[width=0.55\textwidth]{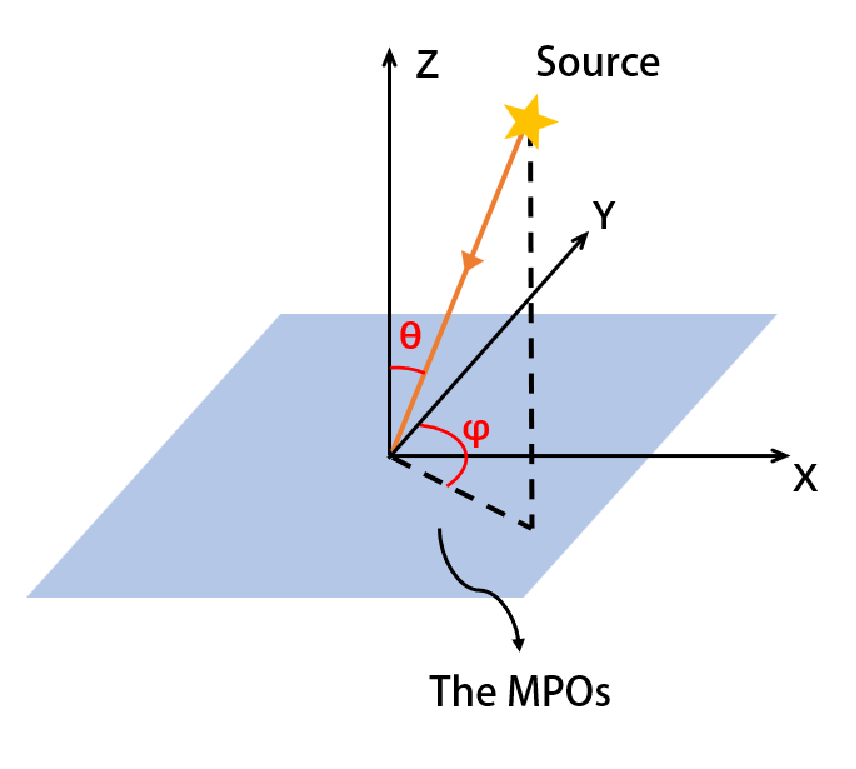}
\caption{Schematic diagram of the relative position relationship between the observed source and the MPOs. In this diagram, the MPOs is approximated as a plane. The plane where the MPOs is located represents the X-Y plane, and the direction perpendicular to the MPOs is the Z-direction. }\label{fig_incident_direction}
\end{figure}

We simulated scenarios in which a contaminating source is present during the observations. All simulations are performed under ideal alignment conditions: the MPOs and the detector system perfectly aligned and the target source is placed on-axis. The spectral index and flux of both target and contaminating sources were assumed to be the same as those of the Crab Nebula. In an observation with an exposure of 200\,s, each source will emit about 1,000,000 photons. To describe the incident direction of the source, we establish a Cartesian coordinate system (see Figure~\ref{fig_incident_direction}) in which the MPOs are approximated as a plane. The surface of the MPOs is defined as the X–Y plane, and the direction normal to the MPOs is taken as the Z axis. Two angles are used to characterize the source direction: off-axis angle $\theta$ is the angle between the incident X-ray photons and the Z axis, and off-axis direction $\phi$ is the angle between the projection of the incident direction onto the X–Y plane and the Y axis. In the X-Y plane, the +Y direction corresponds to $\phi = 0^\circ$, -Y to $\phi = 180^\circ$, +X to $\phi = 90^\circ$, and -X to $\phi = 270^\circ$.

\begin{figure}[h]%
\centering
\includegraphics[width=1.00\textwidth]{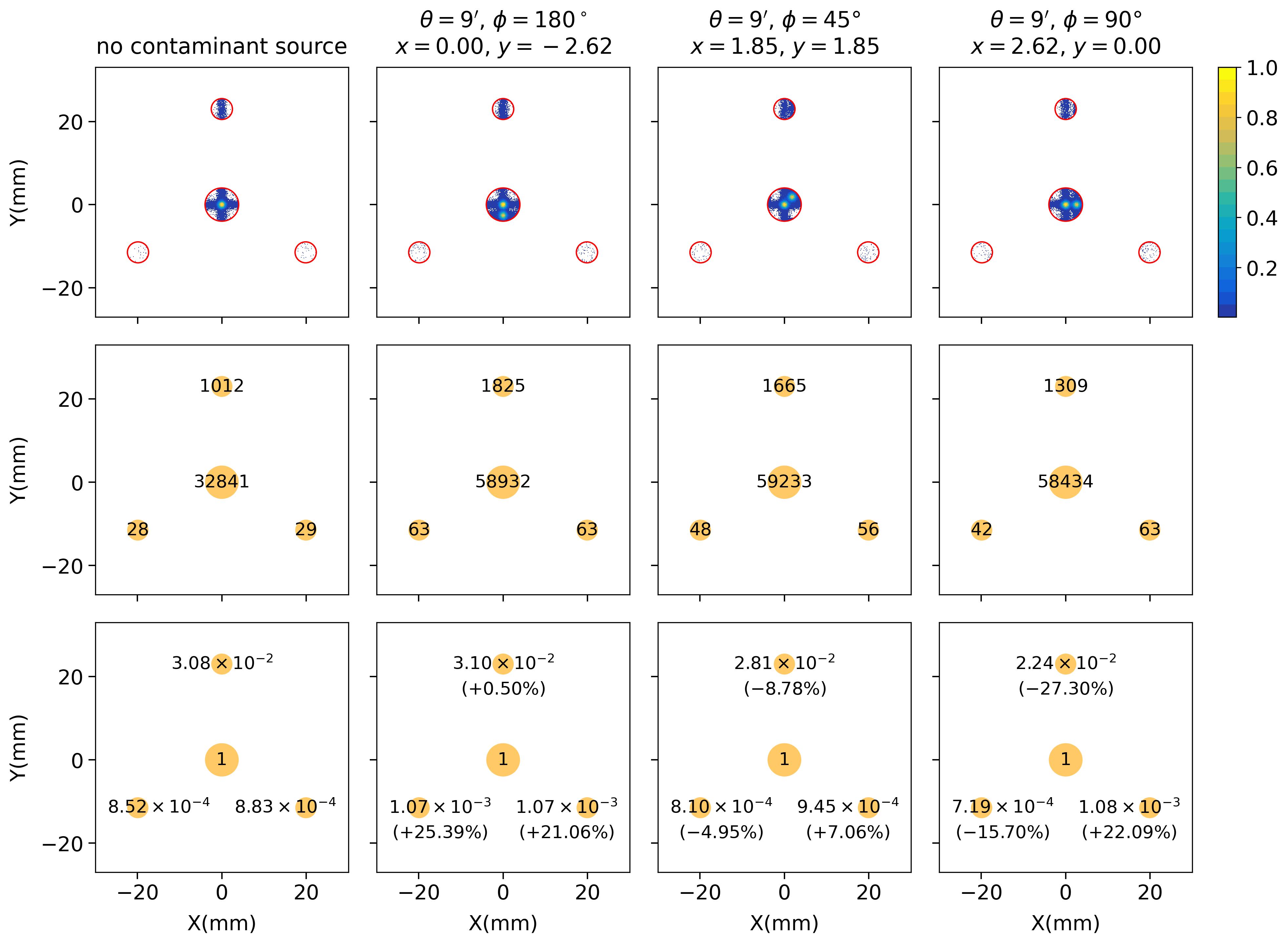}
\caption{The first row shows the PSFs in four different cases at the initial detector layout. The first case does not have the contaminating source and the other three cases have a contaminating source in different positions. The off-axis angle $\theta$ and off-axis direction $\phi$ of the contaminating source, as well as the corresponding focal spot position $(x,y)$ on the focal plane after focusing by the MPOs, are labelled above each panel. The focal plane coordinates are given in units of mm. The second row presents the corresponding count in each detector for different cases. The third row calculates the relative count in each detector. For the three cases with a contaminating source, the changes of the relative counts in the three surrounding detectors compared to the case without contaminating source are also shown. The positive sign indicates an increase and the negative sign indicates a decrease.}\label{fig_compareSDDv1}
\end{figure}

We conducted four sets of comparative experiments, in which the target source was of on-axis incidence, and the contaminating source was of off-axis incidence with an off-axis angle of \ang{;9;}, but with different off-axis directions. The PSFs for each case are shown in the first row of Figure \ref{fig_compareSDDv1}. For the target source, we assume an on-axis incidence, meaning that both the off-axis angle and direction are zero, and the focal spot is located at $(0,0)$ on the focal plane. The position of the contaminating source is defined by its off-axis angle $\theta$ and off-axis direction $\phi$, which are labelled above each case in the Figure \ref{fig_compareSDDv1}. Here, $\theta$ also represents the angular distance to the target source. The focal spot position $(x,y)$ on the focal plane of a source after being focused by the MPOs is related to $(\theta,\phi)$ through:
\begin{equation}
x = f \tan{\theta} \sin{\phi},
\label{eq_x}
\end{equation}
\begin{equation}
y = f \tan{\theta} \cos{\phi},
\label{eq_y}
\end{equation}
where $f = $ 1000\,mm is the focal length of the MPOs. The corresponding focal spot positions of the contaminating source are also indicated in Figure~\ref{fig_compareSDDv1}, with $(x,y)$ expressed in units of millimetres.

Since SDD does not provide position information about the deposited particles and only the count is available, the existence of a contaminating source can only be inferred from the change in count. The second row of Figure \ref{fig_compareSDDv1} displays the simulated count recorded by each detector for the four cases, representing the data available in practical observations. In the last row of Figure \ref{fig_compareSDDv1}, for comparison purposes, the relative count for each detector is calculated. The relative count of the central detector is set to 1, and the relative counts of the three surrounding detectors are the ratio of their counts to that of the central detector, as described in Equation \ref{eq_relative_count},
\begin{equation}
R= \frac{N_{\rm H20}}{N_{\rm H50}}, 
\label{eq_relative_count}
\end{equation}
\begin{equation}
\text{with } \sigma_{R}= \frac{N_{\rm H20}}{N_{\rm H50}}\sqrt{(\frac{\sigma_{N_{\rm H20}}}{N_{\rm H20}})^2+(\frac{\sigma_{N_{\rm H50}}}{N_{\rm H50}})^2}, 
\label{eq_relative_count_sigma}
\end{equation}
where $R$ is the relative count, $N_{\rm H20}$ is the count in one of the three surrounding detectors, and $N_{\rm H50}$ is the count in the central detector. As the counts obtained from simulation include statistical fluctuations and follow a Poisson distribution, their uncertainties are taken as $\sigma_{N_{\rm H20}} = \sqrt{N_{\rm H20}}$ and $\sigma_{N_{\rm H50}} = \sqrt{N_{\rm H50}}$. In order to visually represent changes in these relative counts due to the existence of a contaminating source, we marked the corresponding change below each relative count. The change $C$ is calculated as the percentage difference between the relative counts with and without a contaminating source: 
\begin{equation}
C = \frac{R_{\text{with}} - R_{\text{no}}}{R_{\text{no}}} \times 100\%, 
\label{eq_change}
\end{equation}
\begin{equation}
\text{with } \sigma_{C}= \frac{R_{\text{with}}}{R_{\text{no}}}\sqrt{(\frac{\sigma_{R_{\text{with}}}}{R_{\text{with}}})^2+(\frac{\sigma_{R_{\text{no}}}}{R_{\text{no}}})^2}.
\label{eq_change_sigma}
\end{equation}
In Equation \ref{eq_change}, $R_{\text{with}}$ is the relative count of a surrounding detector with a contaminating source, $R_{\text{no}}$ is the relative count of the same detector without a contaminating source, $\sigma_{R_{\text{with}}}$ and $\sigma_{R_{\text{with}}}$ are derived from Equation~\ref{eq_relative_count_sigma}. The computed values of $C$, $\sigma_{C}$ and their significance levels are summarized in Table \ref{tab_compareSDDv1}. In the subsequent analysis, we consider the change above 5$\sigma$ as a significant change.

\begin{table}[h]
\caption{For the three cases with a contaminating source in Figure \ref{fig_compareSDDv1}, the change in relative count ($C$), standard deviation ($\sigma_{C}$) and significance level of the change ($\lvert$C$\rvert$/$\sigma_{C}$) for each surrounding detector.}
\label{tab_compareSDDv1}%
\begin{tabular}{p{2.5cm}<{\centering}p{1.5cm}<{\centering}p{2.3cm}<{\centering}p{2.3cm}<{\centering}p{2.3cm}<{\centering}}
\toprule
         &      & $C$ & $\sigma_{C}$ & $\lvert$C$\rvert$/$\sigma_{C}$ \\
\midrule
\multirow{3}{*}{$\theta=\ang{;9;}, \phi=\ang{180;;}$} 
        & SDD1 & +0.50\%     & 3.98\%     &   0.12        \\
        & SDD2 & +25.39\%      & 31.00\%    &   0.82          \\
        & SDD3 & +21.06\%      & 29.91\%     &  0.70        \\
\midrule                      
\multirow{3}{*}{$\theta=\ang{;9;}, \phi=\ang{45;;}$} 
        & SDD1 & -8.78\%     & 3.74\%     &   2.34         \\
        & SDD2 & -4.95\%      & 24.72\%    &   0.20          \\
        & SDD3 & +7.06\%      & 26.76\%     &  0.26          \\
\midrule                      
\multirow{3}{*}{$\theta=\ang{;9;}, \phi=\ang{90;;}$} 
        & SDD1 & -27.30\%     & 3.14\%     &   8.69        \\
        & SDD2 & -15.70\%      & 22.82\%    &   0.59          \\
        & SDD3 & +22.09\%      & 30.18\%     &  0.73          \\
\botrule
\end{tabular}
\end{table}

In both the second and third cases, the relative counts in the three surrounding detectors show no significant change compared to the first case which has no contaminating source, making it appear as if there was only one source present. This is because when the count in SDD0 increases due to the existence of a contaminating source inside FOV, the counts in the surrounding three detectors also increase proportionally. In the last case, the contaminating source's horizontal arm overlaps with the target source's horizontal arm, while its vertical arm does not overlap with the target source's vertical arm. As a result, the count in SDD0 nearly doubles, while the count in SDD1, which covers the vertical arm, does not increase much. This causes the relative count in SDD1 to decrease by 27.30$\%$ compared to the case without a contaminating source. Based on this, we can infer the existence of a contaminating source from the horizontal direction.

\subsection{Updated layout}\label{subsec2}

\begin{figure}[h]%
\centering
\includegraphics[width=0.55\textwidth]{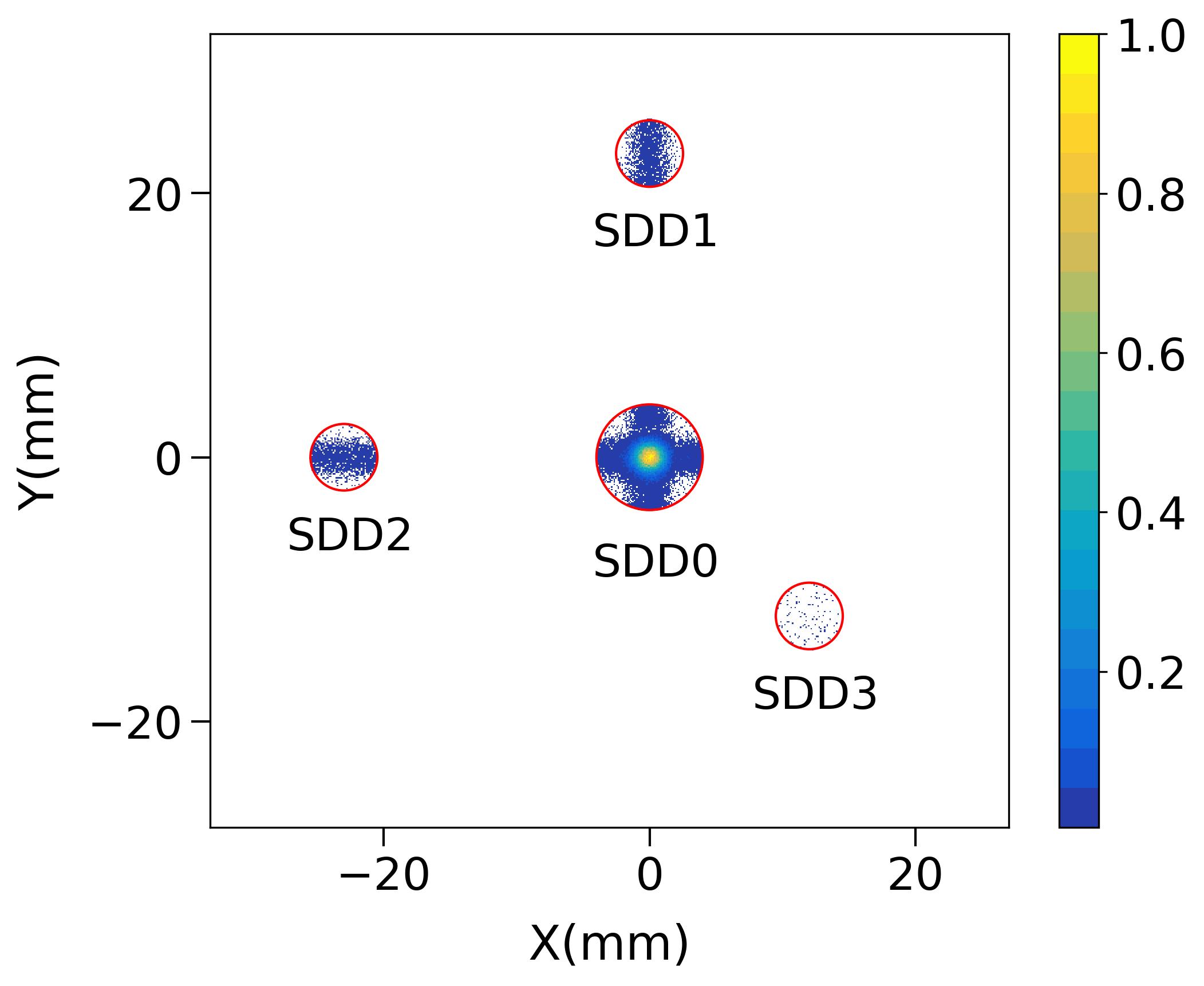}
\caption{PSF in the four detectors for the new detector layout design. }\label{fig_SDDv2psf}
\end{figure}

Drawing inspiration from the fourth case, we explored whether placing a detector on the horizontal arm could allow us to leverage the change in the relative count on it to deduce the existence of a contaminating source from the vertical direction. This led to the proposal of a new detector layout scheme: maintaining the H50 detector at the central position, placing one H20 detector on each of the horizontal and vertical arms, and situating the remaining H20 detector on the diffuse patch to measure the background. In this new design, we adjusted the distance between the background detector and the primary detector. In the original design, the distance between the three surrounding detectors and the central detector was uniformly set at 23 mm. In the new design, the two cross-arm detectors maintain a distance of 23 mm from the central detector, while the distance of the background detector has been reduced to 17 mm. This adjustment was made after carefully considering the geometric dimensions of the detector housing and the background assessment capabilities of the background detector. Specifically, when the background detector is positioned closer to the central detector, the systematic error in evaluating the background is reduced. The PSF of this detector layout is depicted in Figure \ref{fig_SDDv2psf}. The detector on the vertical arm is marked as SDD1, the detector on the horizontal arm as SDD2, and the background detector is SDD3.

\begin{figure}[h]%
\centering
\includegraphics[width=1.00\textwidth]{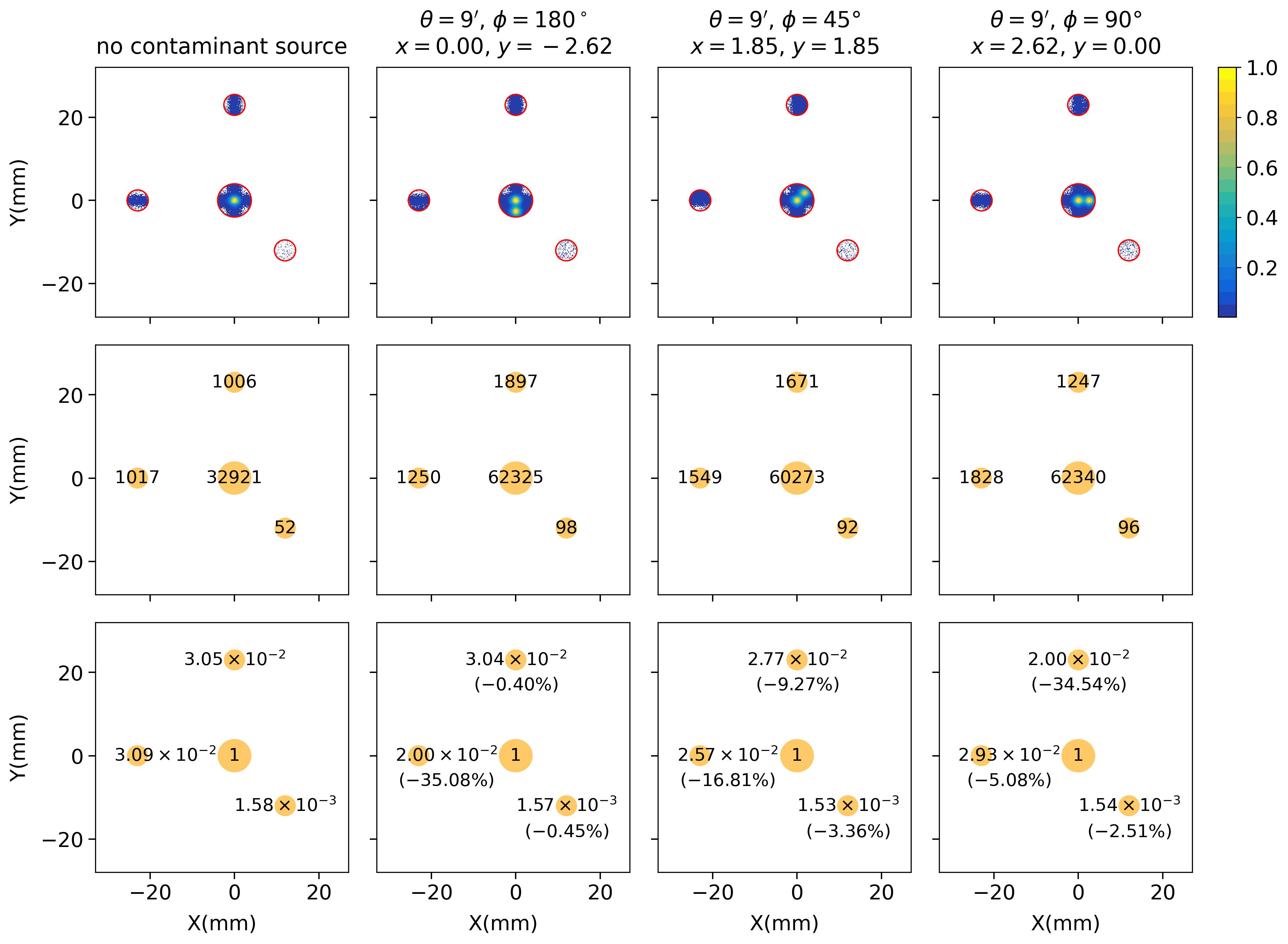}
\caption{Same as Figure \ref{fig_compareSDDv1} but for the new detector layout design.}\label{fig_compareSDDv2}
\end{figure}

We conducted the same four sets of comparison experiments mentioned earlier for this detector layout, as illustrated in Figure \ref{fig_compareSDDv2}. Detailed data related to the changes are listed in Table \ref{tab_compareSDDv2}. In comparison to the first case without a contaminating source, a notable decrease in the relative count of SDD2 is observed in the second case, while a significant decrease in the relative count of SDD1 is evident in the fourth case. These changes enable us to infer the direction of the contaminating source. However, for the third case, we are still unable to ascertain the presence of the contaminating source.

\begin{table}[h]
\caption{Changes in relative count, standard deviation, and significance level of the change for each surrounding detector across three contaminating source cases (refer to Figure \ref{fig_compareSDDv2}). }\label{tab_compareSDDv2}%
\begin{tabular}{p{2.5cm}<{\centering}p{1.5cm}<{\centering}p{2.3cm}<{\centering}p{2.3cm}<{\centering}p{2.3cm}<{\centering}}
\toprule
         &      & $C$ & $\sigma_{C}$ & $\lvert$C$\rvert$/$\sigma_{C}$ \\
\midrule
\multirow{3}{*}{$\theta=\ang{;9;}, \phi=\ang{180;;}$} 
        & SDD1 & -0.40\%     & 3.94\%     &   0.10        \\
        & SDD2 & -35.08\%      & 2.77\%    &   12.64          \\
        & SDD3 & -0.45\%      & 17.68\%     &  0.02         \\
\midrule                      
\multirow{3}{*}{$\theta=\ang{;9;}, \phi=\ang{45;;}$} 
        & SDD1 & -9.27\%     & 3.70\%     &   2.50         \\
        & SDD2 & -16.81\%      & 3.41\%    &   4.92          \\
        & SDD3 & -3.36\%      & 17.42\%     &  0.19          \\
\midrule                      
\multirow{3}{*}{$\theta=\ang{;9;}, \phi=\ang{90;;}$} 
        & SDD1 & -34.54\%     & 2.81\%     &   12.29         \\
        & SDD2 & -5.08\%      & 3.77\%    &   1.35          \\
        & SDD3 & -2.51\%      & 17.47\%     &  0.14          \\
\botrule
\end{tabular}
\end{table}

\begin{figure}[h]%
\centering
\includegraphics[width=0.80\textwidth]{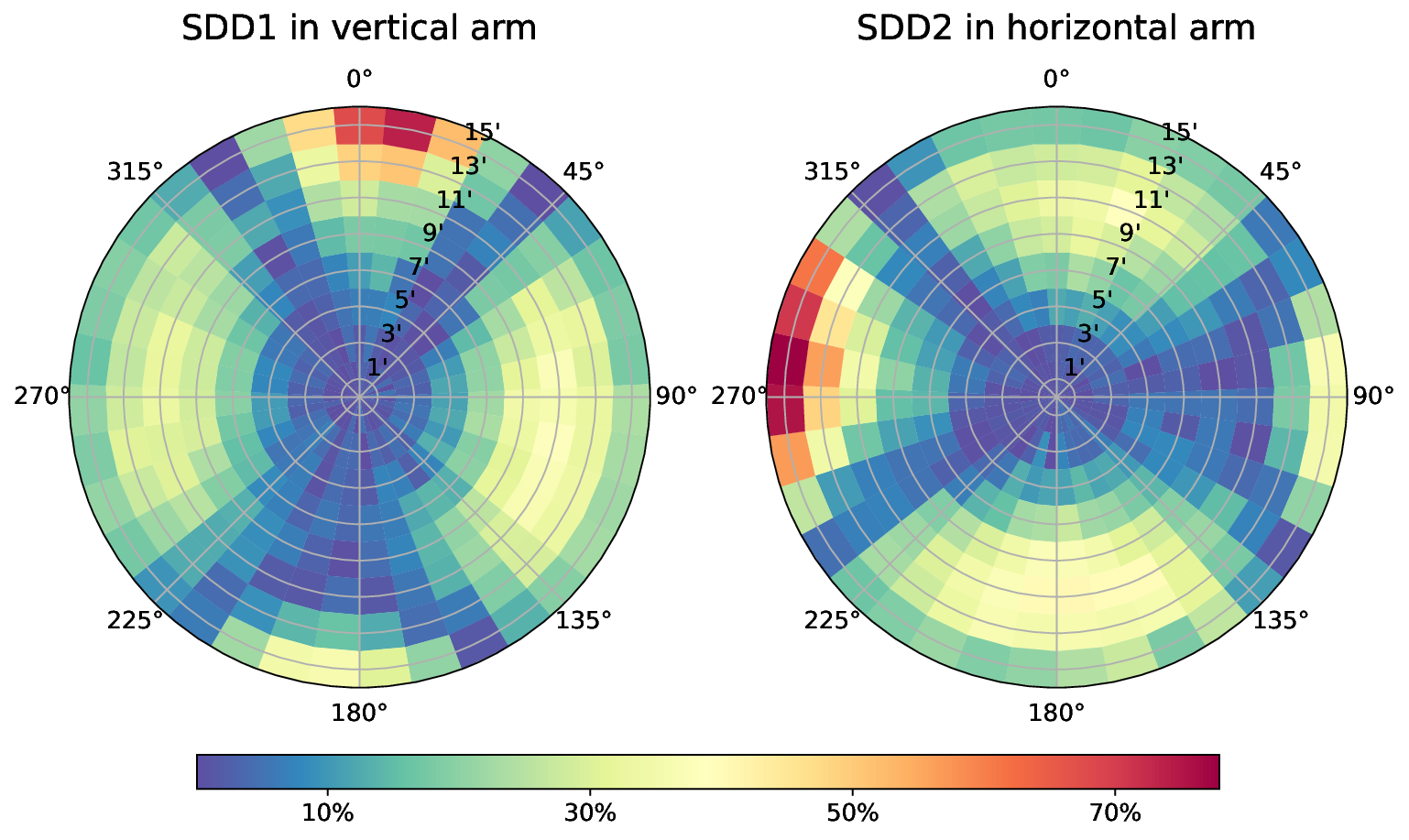}
\caption{Relative-count variations for SDD1 (left panel) and SDD2 (right panel) as a contaminating source moves across off-axis angles $\theta=$\ang{;0;}--\ang{;15;} (increasing outward).}\label{fig_contaminant}
\end{figure}
   
We simulated cases in which the contaminating source was positioned at various off-axis angles and directions, ranging from \ang{0;;} to \ang{360;;} for off-axis direction and from \ang{;0;} to \ang{;15;} for off-axis angle. The changes of the relative counts in the two cross-arms detectors are shown in Figure \ref{fig_contaminant}. Here, the polar radius represents the off-axis angle, and the polar angle represents the off-axis direction. The color bar indicates the variations in relative counts. For SDD1, there are four regions that show relatively large changes (corresponding to brighter colors), where the off-axis direction is around \ang{90;;} and \ang{270;;} with the off-axis angle between \ang{;8;} and $\ang{;13;}$, and the off-axis direction around \ang{0;;} and \ang{180;;} with the off-axis angle greater than \ang{;13;}. For SDD2, a similar pattern is observed, with four regions of notable change, where the off-axis direction is around \ang{0;;} and \ang{180;;} with the off-axis angle between \ang{;8;} and \ang{;13;}, and the off-axis direction is around \ang{90;;} and \ang{270;;} with the off-axis angle greater than \ang{;13;}.

\begin{figure}[h]%
\centering
\includegraphics[width=1.00\textwidth]{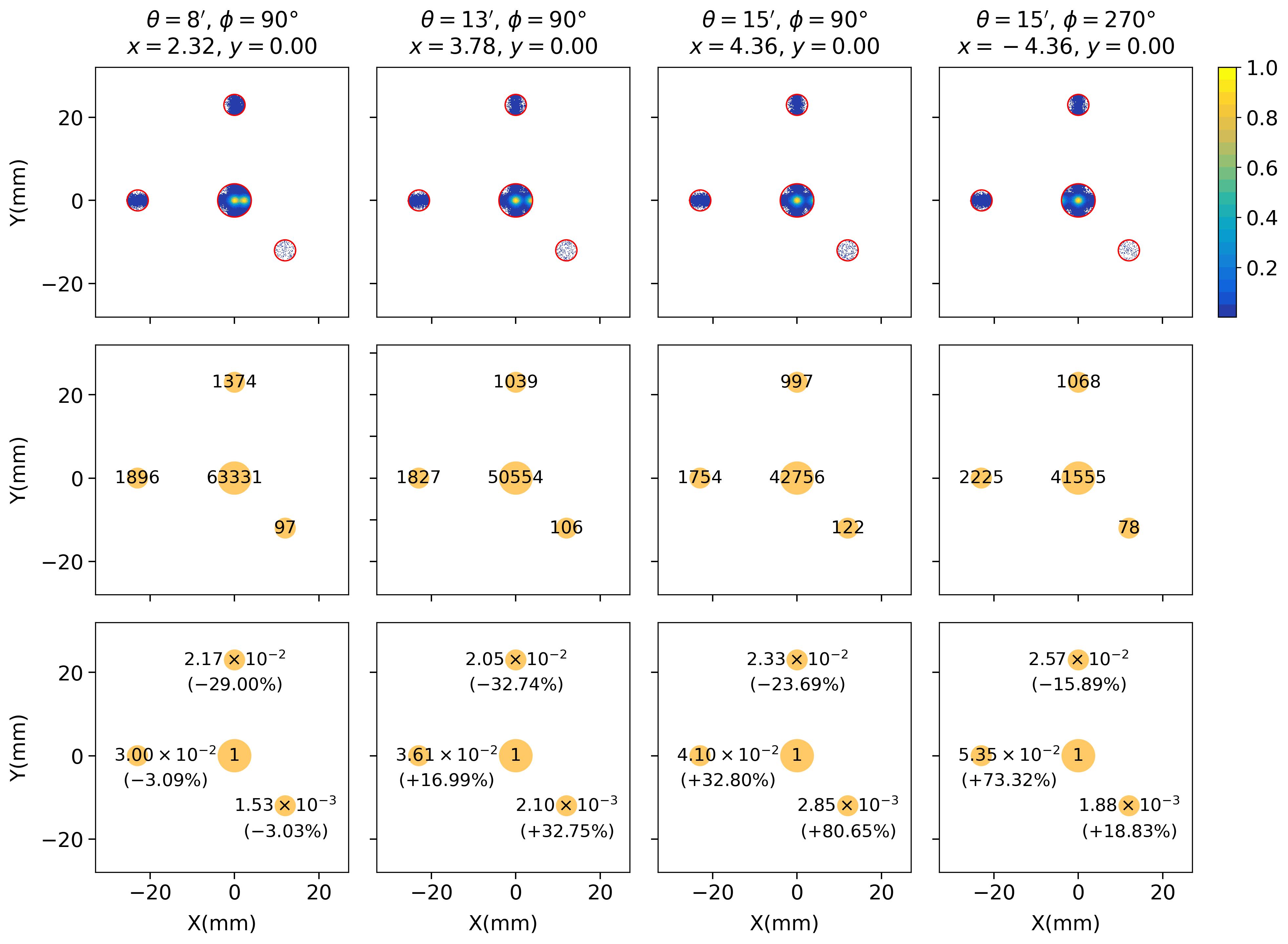}
\caption{PSF, count, relative count, and relative count change for each detector in the new layout with contaminating sources from specific off-axis angles in the horizontal direction. }\label{fig_explainSDDv2}
\end{figure}

We will explain the results by taking the example of the contaminating source located at the horizontal direction, i.e., the off-axis direction is \ang{90;;} or \ang{270;;}. When the off-axis angle of the contaminating source reaches \ang{;8;}, its focal spot is within SDD0 while a considerable portion of the contaminating source's vertical arm has moved out of SDD1, causing a noticeable decrease in the relative count of SDD1, as shown in the first case in Figure \ref{fig_explainSDDv2}. As the off-axis angle progresses from \ang{;8;} to \ang{;13;}, the vertical arm of the contaminating source continues to move outward, while the majority of the contaminating source's focal spot remains within SDD0, leading to a sustained large change in the relative count of SDD1. The second case in Figure \ref{fig_explainSDDv2} corresponds to the off-axis angle of \ang{;13;}. When the off-axis angle is larger than \ang{;13;}, although the vertical arm of the contaminating source has completely exited SDD1, its focal spot has also gradually moved out of SDD0. Consequently, the decrease in the relative count of SDD1 becomes smaller. However, due to the superimposition of the horizontal arms of the target source and the contaminating source in SDD2, the relative count of SDD2 will increase when the off-axis angle exceeds \ang{;13;}. The third case in Figure \ref{fig_explainSDDv2} corresponds to the off-axis angle reaching \ang{;15;}, where the decrease in the relative count of SDD1 is reduced to 23.69$\%$, while SDD2's relative count shows a 32.80$\%$ increase. It is also interesting to note that in SDD2, for the contaminating source with an off-axis angle larger than \ang{;13;} and coming from the horizontal direction, the contaminating source from the \ang{270;;} direction results in a more pronounced change in the relative count compared to contaminating source from the \ang{90;;} direction. This is because the closer the cross-arms are to the central focal spot, the higher the counts. At an off-axis angle of \ang{;15;}, the count in SDD2 for the contaminating source from the \ang{270;;} direction is larger than that from the \ang{90;;} direction, resulting in a larger increase in the relative count in SDD2. The fourth case in Figure \ref{fig_explainSDDv2} demonstrates the contaminating source at an off-axis angle of \ang{;15;} and an off-axis direction of \ang{270;;}, the relative count in SDD2 increases by 73.32$\%$. Table \ref{tab_explainSDDv2} provides detailed data on the changes in each of the four cases shown in Figure \ref{fig_explainSDDv2}.

\begin{table}[h]
\caption{Changes in relative count, standard deviation, and significance level of the change for each surrounding detector across the four cases (refer to Figure \ref{fig_explainSDDv2}).}\label{tab_explainSDDv2}%
\begin{tabular}{p{2.5cm}<{\centering}p{1.5cm}<{\centering}p{2.3cm}<{\centering}p{2.3cm}<{\centering}p{2.3cm}<{\centering}}
\toprule
         &      & $C$ & $\sigma_{C}$ & $\lvert$C$\rvert$/$\sigma_{C}$ \\
\midrule
\multirow{3}{*}{$\theta=\ang{;8;}, \phi=\ang{90;;}$} 
        & SDD1 & -29.00\%     & 3.02\%     &   9.61         \\
        & SDD2 & -3.09\%      & 3.84\%    &   0.80          \\
        & SDD3 & -3.03\%      & 17.52\%     &  0.17          \\
\midrule                      
\multirow{3}{*}{$\theta=\ang{;13;}, \phi=\ang{90;;}$} 
        & SDD1 & -32.74\%     & 3.05\%     &   10.74         \\
        & SDD2 & +16.99\%      & 4.68\%    &   3.63          \\
        & SDD3 & +32.75\%      & 23.63\%     &  1.38          \\
\midrule                      
\multirow{3}{*}{$\theta=\ang{;15;}, \phi=\ang{90;;}$} 
        & SDD1 & -23.69\%     & 3.47\%     &   6.83         \\
        & SDD2 & +32.80\%      & 5.41\%    &   6.06          \\
        & SDD3 & +80.65\%      & 31.64\%     &  2.55          \\
        \midrule                      
\multirow{3}{*}{$\theta=\ang{;15;}, \phi=\ang{270;;}$} 
        & SDD1 & -15.89\%     & 3.78\%     &   4.20         \\
        & SDD2 & 73.32\%      & 6.80\%    &   10.78          \\
        & SDD3 & 18.83\%      & 22.39\%     &  0.84          \\
\botrule
\end{tabular}
\end{table}

\begin{figure}[h]%
\centering
\includegraphics[width=0.45\textwidth]{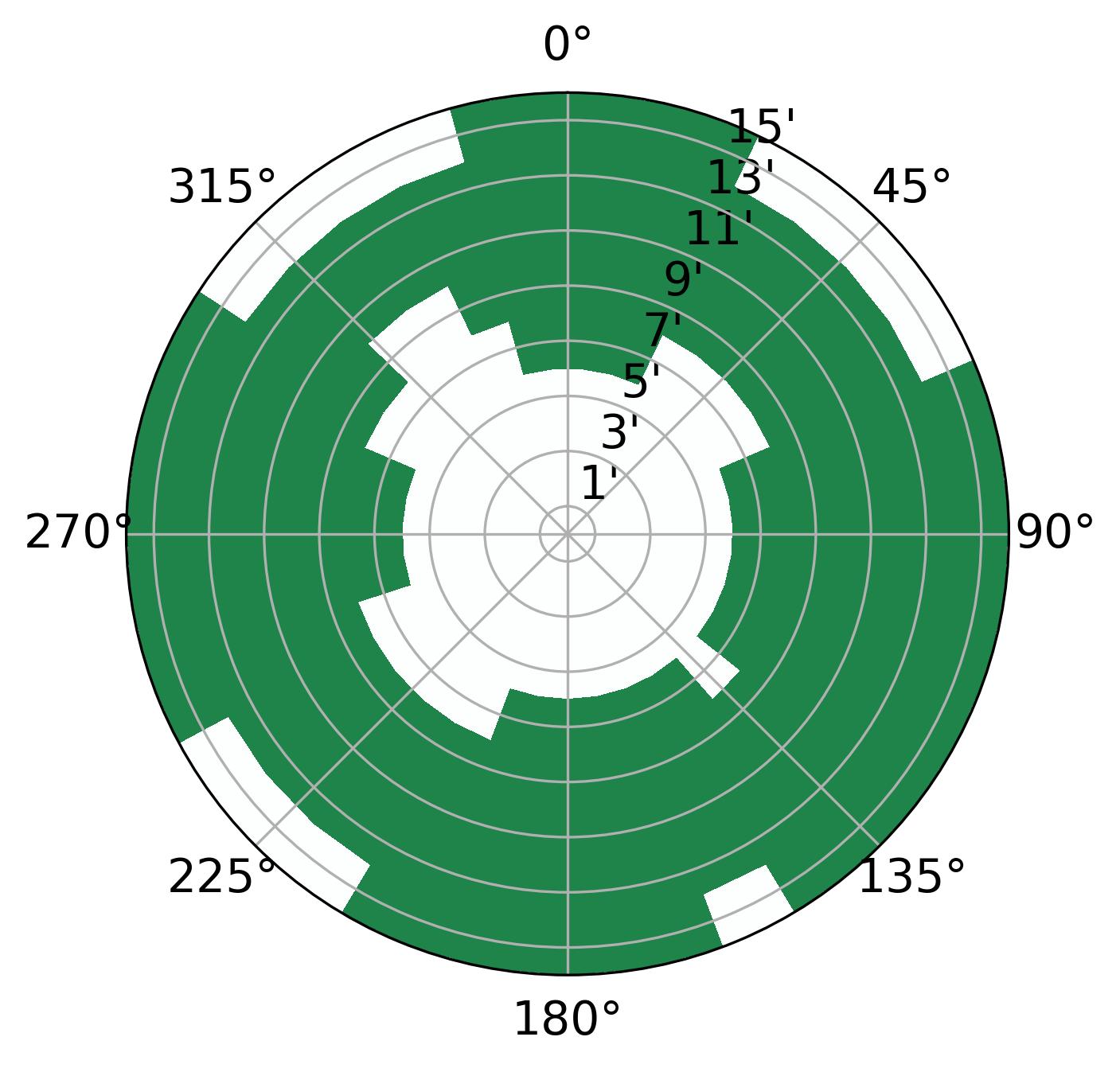}
\caption{Grids colored in green in this polar coordinate indicate that contaminating sources from these off-axis angles and off-axis directions can cause the change of the relative count in either of the two cross-arms detectors to exceed 5$\sigma$.}\label{fig_contaminant5sigma}
\end{figure}

We plotted the cases where the changes exceeded 5$\sigma$ in the two cross-arms detectors in one polar coordinate, as shown in Figure \ref{fig_contaminant5sigma}. It can be seen that when the off-axis angle of the contaminating source exceeds $\ang{;8;}$, the detector system in this layout can effectively identify the existence of the contaminating source. Contaminating sources from directions around $\ang{0;;}$, $\ang{90;;}$, $\ang{180;;}$ or $\ang{270;;}$ are more easily discernible compared to those from $\ang{45;;}$, $\ang{135;;}$, $\ang{225;;}$ or $\ang{315;;}$. The slight asymmetries on either side of the $\ang{315;;}$-$\ang{135;;}$ symmetry axis are caused by statistical fluctuations. In the following text, the pictures depicting cases where the changes exceeded 5$\sigma$ in the two cross-arms detectors in one polar coordinate also exhibit similar statistical fluctuations. When the off-axis angle of the contaminating source exceeds $\ang{;15;}$, only those from near the horizontal or vertical directions can be detected. As the off-axis angle gets larger, the increase in relative counts of the cross-arms detectors gets smaller and smaller, making the contaminating source less recognizable. However, the impact of the contaminating source on the target source in the primary detector also decreases. For a contaminating source positioned at an off-axis direction of $\ang{90;;}$, when its off-axis angle reaches $0.01^{\circ}/\mathrm{s}$, the increase in relative counts on SDD2 drops below 5$\sigma$, and the count from the contaminating source contributes less than 5$\%$ in the primary detector.

\subsection{Discussion}\label{subsec3}

During the observation, the count recorded by each detector includes counts from the source and the background. The relative counts described above are obtained by directly dividing the total counts of the cross-arms detectors by the total counts of the primary detector. This is reasonable for a strong source with a flux of 1\,Crab, where the background counts in the primary detector and two cross-arms detectors are negligible, accounting for less than 0.1\% of the total counts. Additionally, the background component in the background detector is less than 5\%, making it inappropriate to simply regard the count in the background detector as being all from the background. However, when observing weak sources, such as one with a flux of 1\,mCrab, demanding a 1000-fold increase in exposure time to accumulate the same source counts as a 1\,Crab-flux target. As the exposure time increases, the background count also rises. In this situation, the background count in the background detector will dominate, reaching over 97\% of the total count, indicating that the background detector effectively serves its purpose of measuring the background and its count can be considered as the background count. Meanwhile, the background contributions in the primary detector and cross-arms detectors become significant as well, accounting for 10\% and 58\%, respectively. Therefore, the background detector's count can be used to subtract the corresponding background count from the primary and cross-arm detectors, according to their inter-detector background ratios. After this subtraction, further analysis can be conducted, including calculating the relative counts and changes for the two cross-arms detectors. 

\begin{figure}[h]%
\centering
\includegraphics[width=0.8\textwidth]{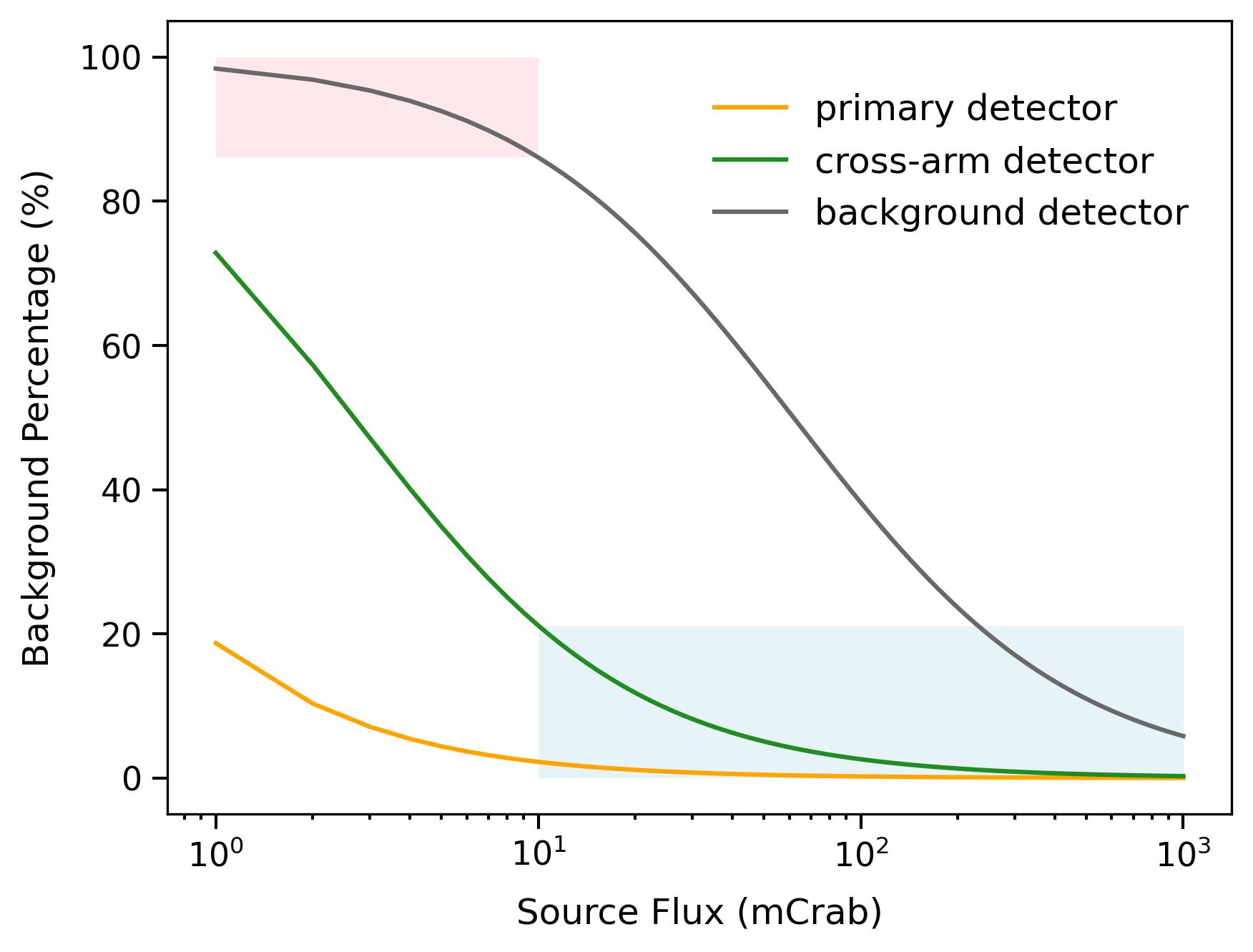}
\caption{Background fraction versus source flux for the three types of detectors.}\label{fig_bkg_percentage}
\end{figure} 

\begin{figure}[h]
\centering
    \begin{minipage}{0.315\textwidth}
        \includegraphics[width=\textwidth]{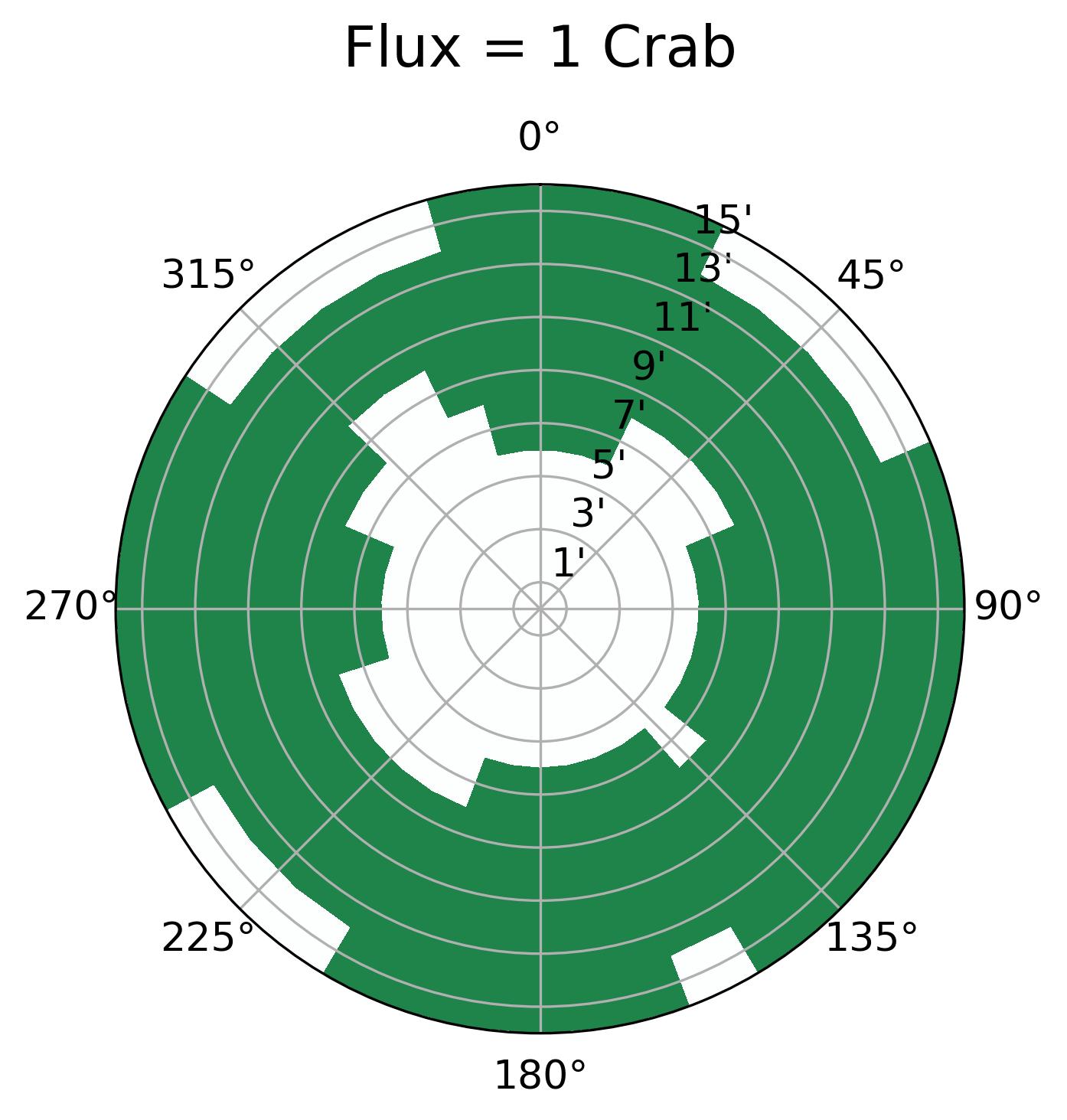}
    \end{minipage}
    \hspace{0.05\textwidth}
    \begin{minipage}{0.315\textwidth}
        \includegraphics[width=\textwidth]{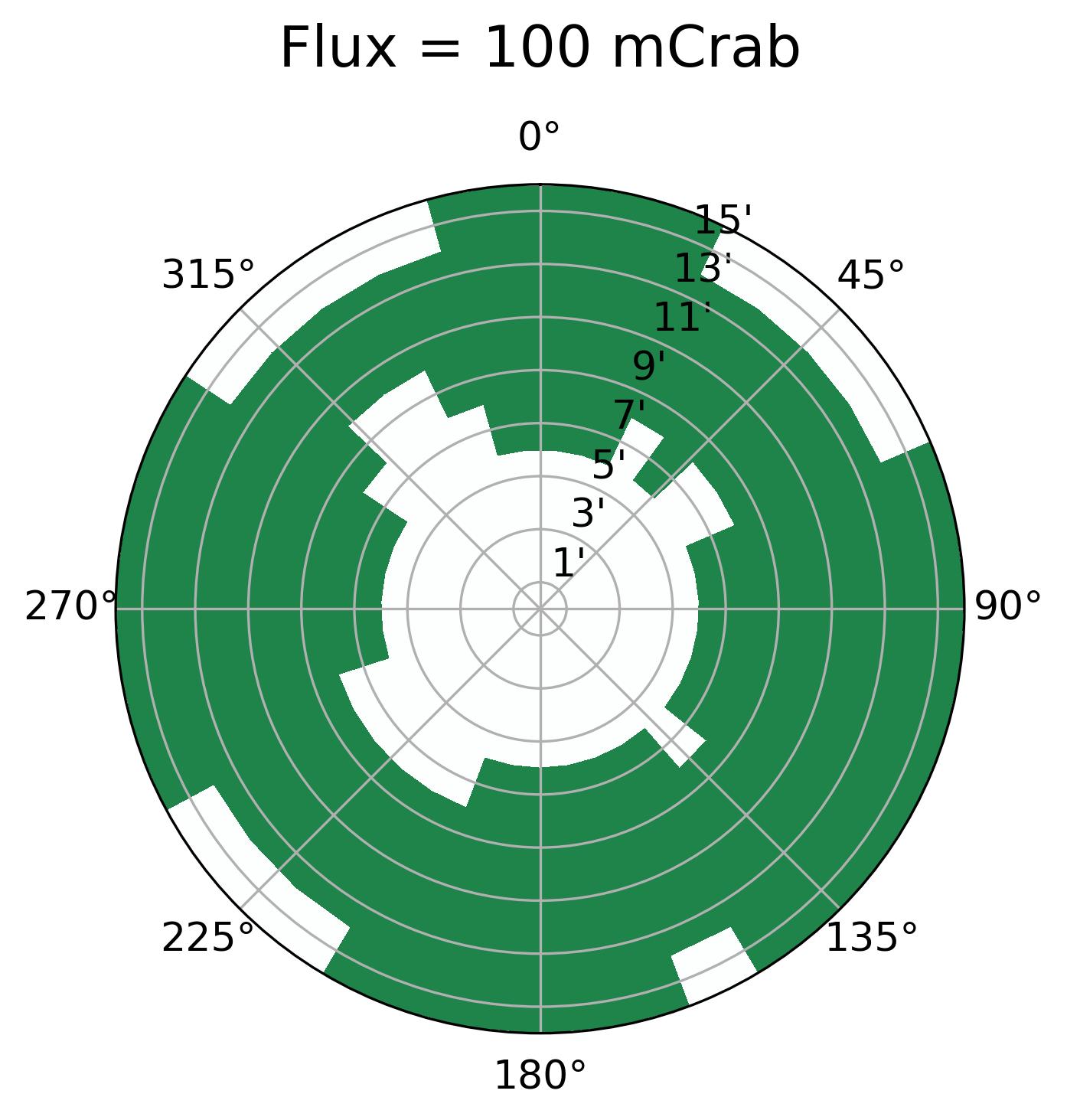}
    \end{minipage}
    \vspace{2em} 
    \begin{minipage}{0.315\textwidth}
        \includegraphics[width=\textwidth]{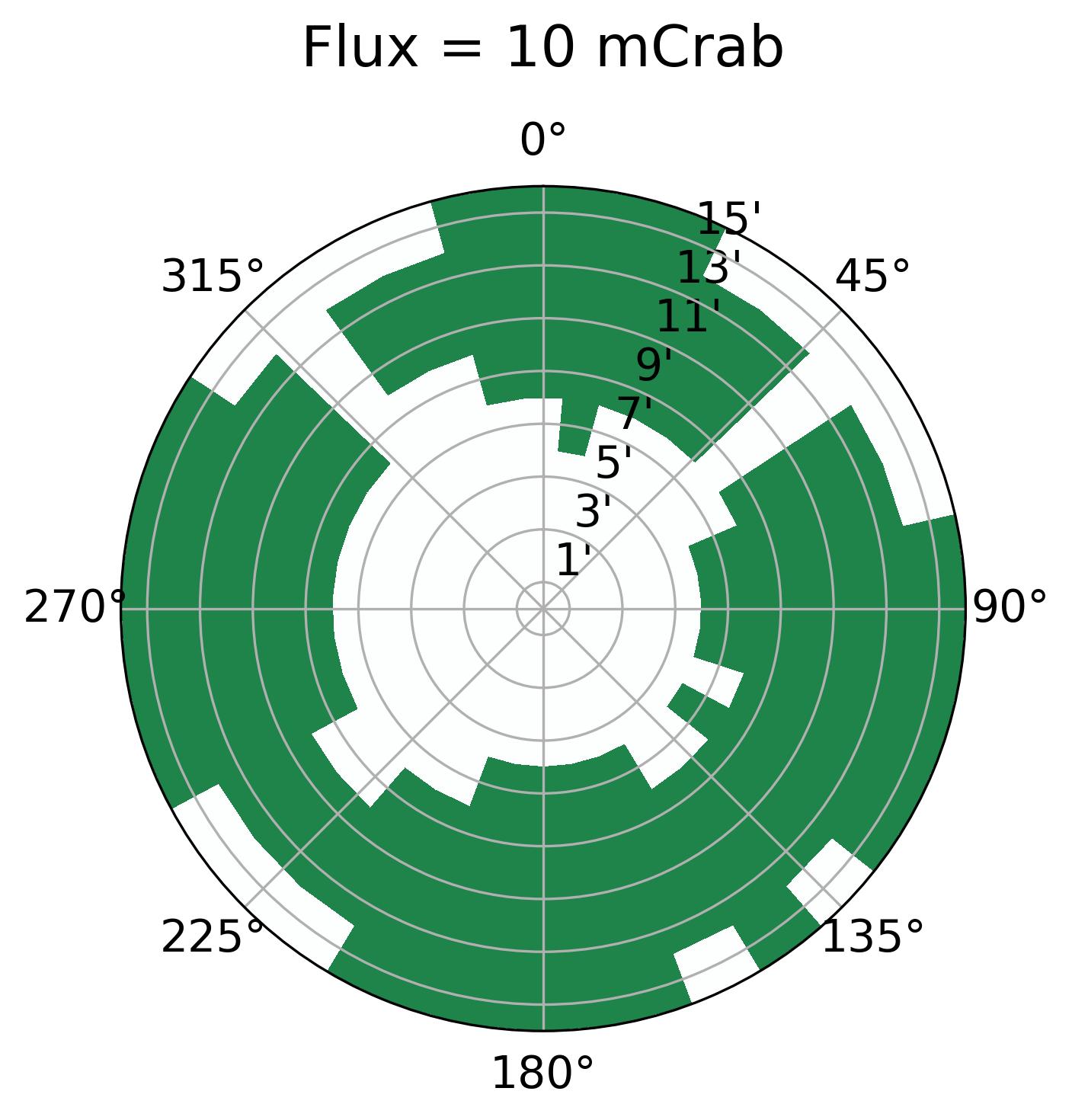}
    \end{minipage}
    \hspace{0.05\textwidth}
    \begin{minipage}{0.315\textwidth}
        \includegraphics[width=\textwidth]{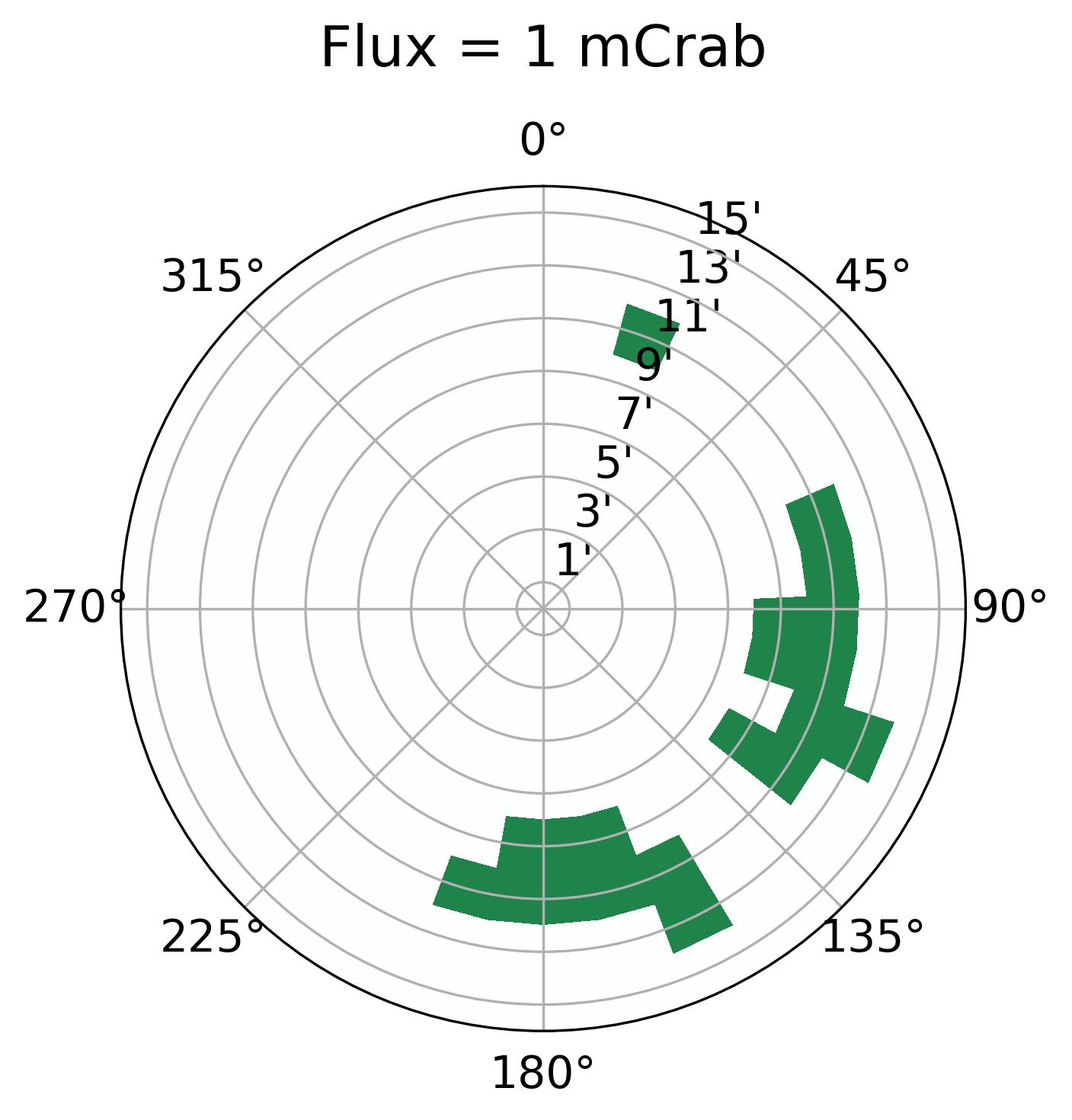}
    \end{minipage}
    \caption{Cases in which the variation in either cross-arm detector exceeds 5$\sigma$, shown for four flux levels (indicated above each panel). The target and contaminating sources have equal flux. Background counts are retained for the top-row cases and subtracted for those in the bottom row.}
    \label{fig_contaminantflux}
\end{figure}

To decide whether background subtraction is required, we examined the background fraction in each of the three detector types as a function of source flux (Figure~\ref{fig_bkg_percentage}). For every flux, the exposure time is scaled so that the source delivers the same number of counts as a 1-Crab flux source observed for 1000\,s; weaker sources therefore accumulate proportionally longer exposures and higher background levels. Based on this variation, the threshold for background subtraction is set at 10\,mCrab for the following reasons. When the flux of the observed source exceeds 10\,mCrab, the background contributions in both the primary detector and cross-arms detectors are less than 20\%, as depicted by the blue shading in Figure \ref{fig_bkg_percentage}, and it is deemed unnecessary to subtract the background count during data processing. In contrast, when the flux is below 10\,mCrab, the background contribution in the background detector exceeds 85\%, represented by the pink shading in Figure \ref{fig_bkg_percentage}. Consequently, the entire count recorded by the background detector can be treated as background and used to scale-subtract the corresponding background from the primary and cross-arm detectors. Figure~\ref{fig_contaminantflux} shows how the Pathfinder’s ability to identify contaminating sources changes when the target and the contaminating are observed at four flux levels. Performance at 100 mCrab is essentially identical to that at 1\,Crab; it drops markedly as the flux decreases to 10\,mCrab and falls further at 1\,mCrab.

\begin{figure}[h]
    \centering
    \begin{minipage}{0.315\textwidth}
        \includegraphics[width=\textwidth]{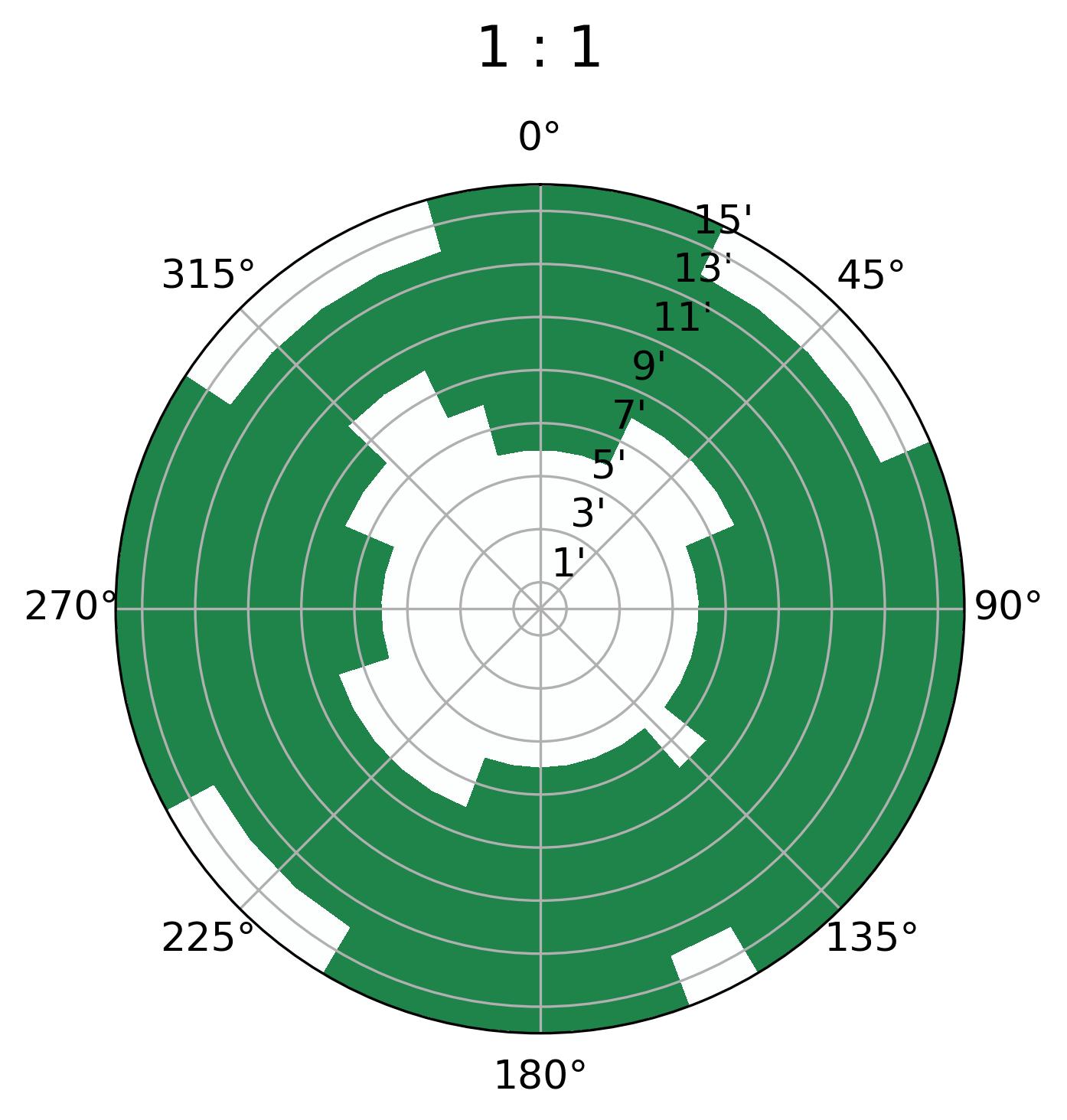}
    \end{minipage}
    \hfill
    \begin{minipage}{0.315\textwidth}
        \includegraphics[width=\textwidth]{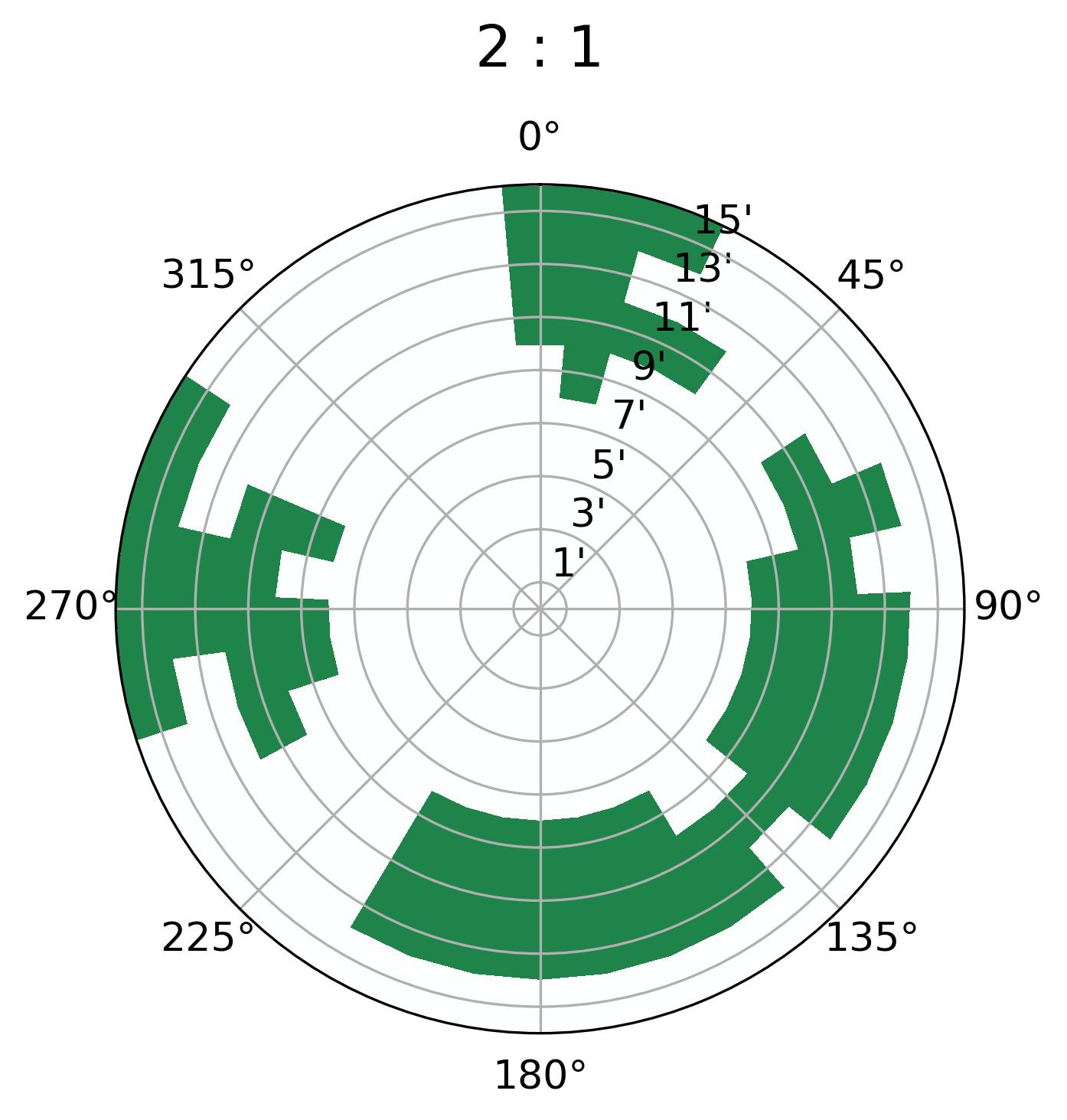}
    \end{minipage}
    \hfill
    \begin{minipage}{0.315\textwidth}
        \includegraphics[width=\textwidth]{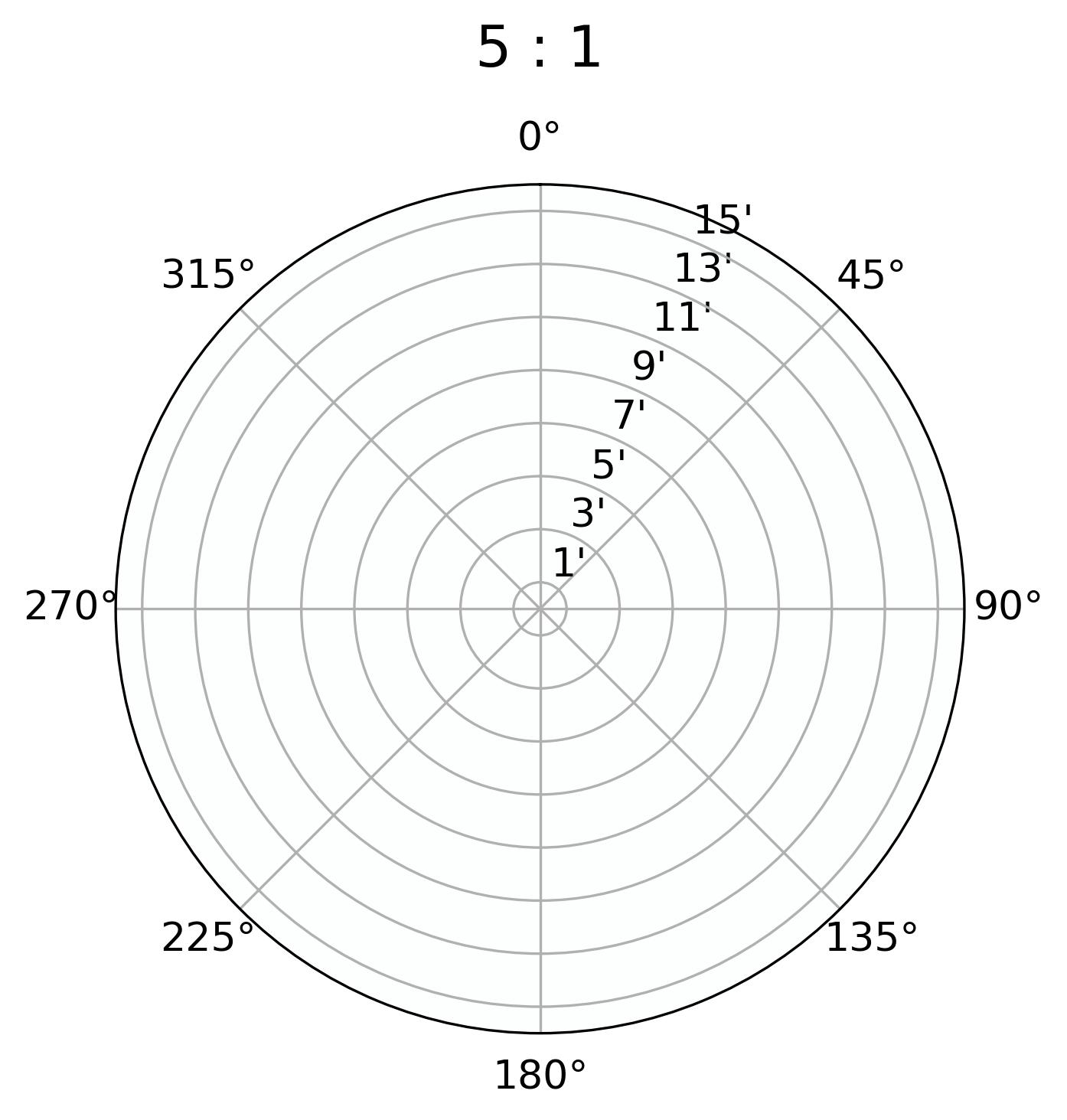}
    \end{minipage}
    \caption{Cases in which the variation in either cross-arm detector exceeds 5$\sigma$ for three target-to-contamination flux ratios (indicated above each panel). In all cases the target source is fixed at a flux of 1\,Crab.}
    \label{fig_contaminantratio}
\end{figure}

Detection performance hinges on the target-to-contamination flux ratio. Figure \ref{fig_contaminantratio} illustrates the Pathfinder’s ability to identify contaminating sources for three ratios. As the ratio declines, contamination identification weakens, yet the contamination’s impact on the target source also lessens.

The results are independent of the source spectrum. Table~\ref{table_spectra} lists the relative counts recorded by the three surrounding detectors for on-axis sources with three representative spectra: a Crab-like power law, and the typical soft and hard spectra of black-hole binaries (BHBs) \citep{bhb}---the primary science targets of the CATCH Type-A Pathfinder. Variations between spectral types fall well within statistical fluctuations, so contaminating source identification is equally effective across all spectra considered.

\begin{table}[h]
\renewcommand\arraystretch{1.5}
\caption{Relative counts in the three surrounding detectors for on-axis sources of various spectral types.}\label{table_spectra}%
\begin{tabular}{p{3.0cm}<{\centering}p{2.3cm}<{\centering}p{2.3cm}<{\centering}p{2.3cm}<{\centering}}
\toprule
\multirow{2}{*}{Spectra} & \multicolumn{3}{@{}c@{}}{Relative counts} \\
\cmidrule{2-4}%
    &  SDD1 & SDD2 & SDD3  \\
\midrule
Crab    &  $3.05\times10^{-2}$ & $3.09\times10^{-2}$ & $1.58\times10^{-3}$  \\
Soft state BHBs &  $3.10\times10^{-2}$ & $3.08\times10^{-2}$ & $1.53\times10^{-3}$ \\
Hard state BHBs & $3.05\times10^{-2}$ & $3.04\times10^{-2}$ & $1.56\times10^{-3}$  \\
\botrule
\end{tabular}
\end{table}

In summary, the cross-arm detectors serve as an efficient and essential tool for contaminating source identification, where monitoring their relative count variations enables reliable detection of the presence of a contaminating source.

\section{Target source position}\label{sec5}

The FOV of the Pathfinder of CATCH Type-A is $\ang{0.4;;}\times\ang{0.4;;}$, indicating that the focal spot cannot be completely detected by the primary detector when the off-axis angle of the target source exceeds \ang{;12;}. With only the primary detector in use, the satellite's capability is limited to distinguishing whether the target source is within or outside the FOV. The second detector layout proposed in Section \ref{sec4} leverages the distinctive cross-shaped PSF unique to MPOs and the changes in relative counts for detectors at specific focal plane positions, which not only allows the identification of contaminating sources, but also improves the positioning accuracy.

\begin{figure}[h]%
\centering
\includegraphics[width=1.00\textwidth]{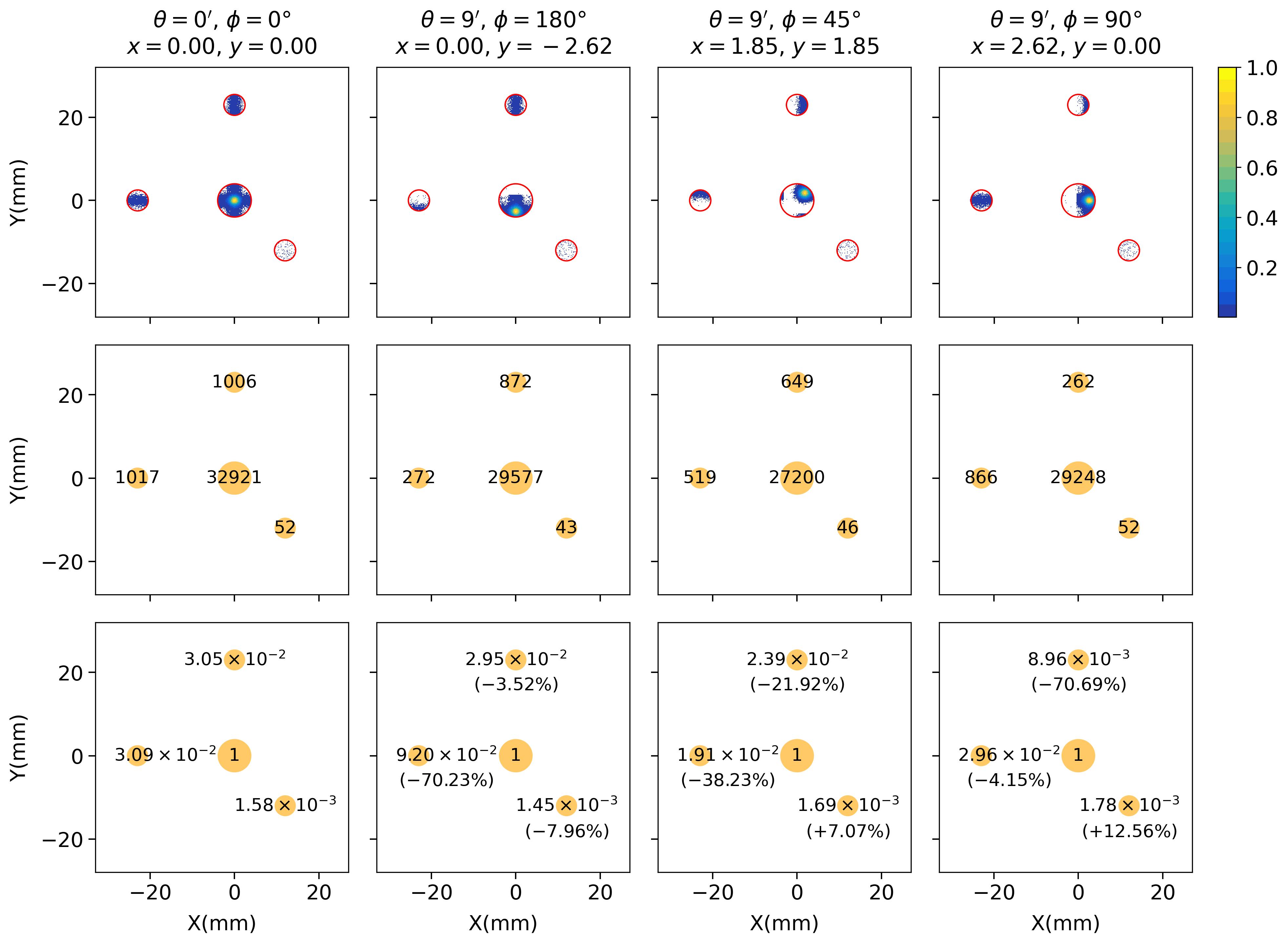}
\caption{PSF, total counts, and relative counts in each detector are shown for the target source arriving from four directions. The off-axis angle $\theta$ and off-axis direction $\phi$, together with the corresponding focal spot position $(x,y)$ on the focal plane after focusing by the MPOs, are labelled above each panel. The focal plane coordinates are given in units of mm. For the three off-axis cases, the panels also display the changes in relative count for each surrounding detector relative to the on-axis case.}\label{fig_positionSDDv2}
\end{figure}

Initially, we conducted simulated observations for the target source from four different positions within the FOV. In the simulation, the target source is assumed to be a Crab-like source with a flux of 1\,Crab. The target source generates 1,000,000 particles in each observation, corresponding to an exposure time of approximately 200\,s. In the first simulation, the target source is on-axis incidence, and in the other three simulations, the target source is off-axis incident with an off-axis angle of $\ang{;9;}$, but with different off-axis directions. In Figure \ref{fig_positionSDDv2}, the first row displays their PSFs, the second row shows the count in each detector, and the third row presents the relative count. Table \ref{tab_positionSDDv2} lists the detailed data related to the changes for the latter three cases. It is evident that all three off-axis incidences can make a significant change in the relative counts on the cross-arms detectors compared to the on-axis incidence.

\begin{table}[h]
\caption{The change in relative count, standard deviation, and significance level of the change for each surrounding detector across the four cases (refer to Figure \ref{fig_positionSDDv2}). }
\label{tab_positionSDDv2}%
\begin{tabular}{p{2.5cm}<{\centering}p{1.5cm}<{\centering}p{2.3cm}<{\centering}p{2.3cm}<{\centering}p{2.3cm}<{\centering}}
\toprule
         &      & $C$ & $\sigma_{C}$ & $\lvert$C$\rvert$/$\sigma_{C}$ \\
\midrule
\multirow{3}{*}{$\theta=\ang{;9;}, \phi=\ang{180;;}$} 
        & SDD1 & -3.52\%     & 4.51\%     &   0.78         \\
        & SDD2 & -70.23\%      & 2.07\%    &   33.87          \\
        & SDD3 &  -7.96\%      & 19.78\%     &  0.40          \\
\midrule                      
\multirow{3}{*}{$\theta=\ang{;9;}, \phi=\ang{45;;}$} 
        & SDD1 & -21.92\%     & 3.96\%     &   5.54         \\
        & SDD2 & -38.23\%      & 3.40\%    &   11.25          \\
        & SDD3 & +7.07\%      & 22.85\%     &  0.31          \\
\midrule                      
\multirow{3}{*}{$\theta=\ang{;9;}, \phi=\ang{90;;}$} 
        & SDD1 & -70.69\%     & 2.05\%     &   34.45         \\
        & SDD2 & -4.15\%      & 4.47\%    &   0.93          \\
        & SDD3 & +12.56\%      & 23.45\%     &  0.53          \\
\botrule
\end{tabular}
\end{table}

\begin{figure}[h]%
\centering
\includegraphics[width=0.80\textwidth]{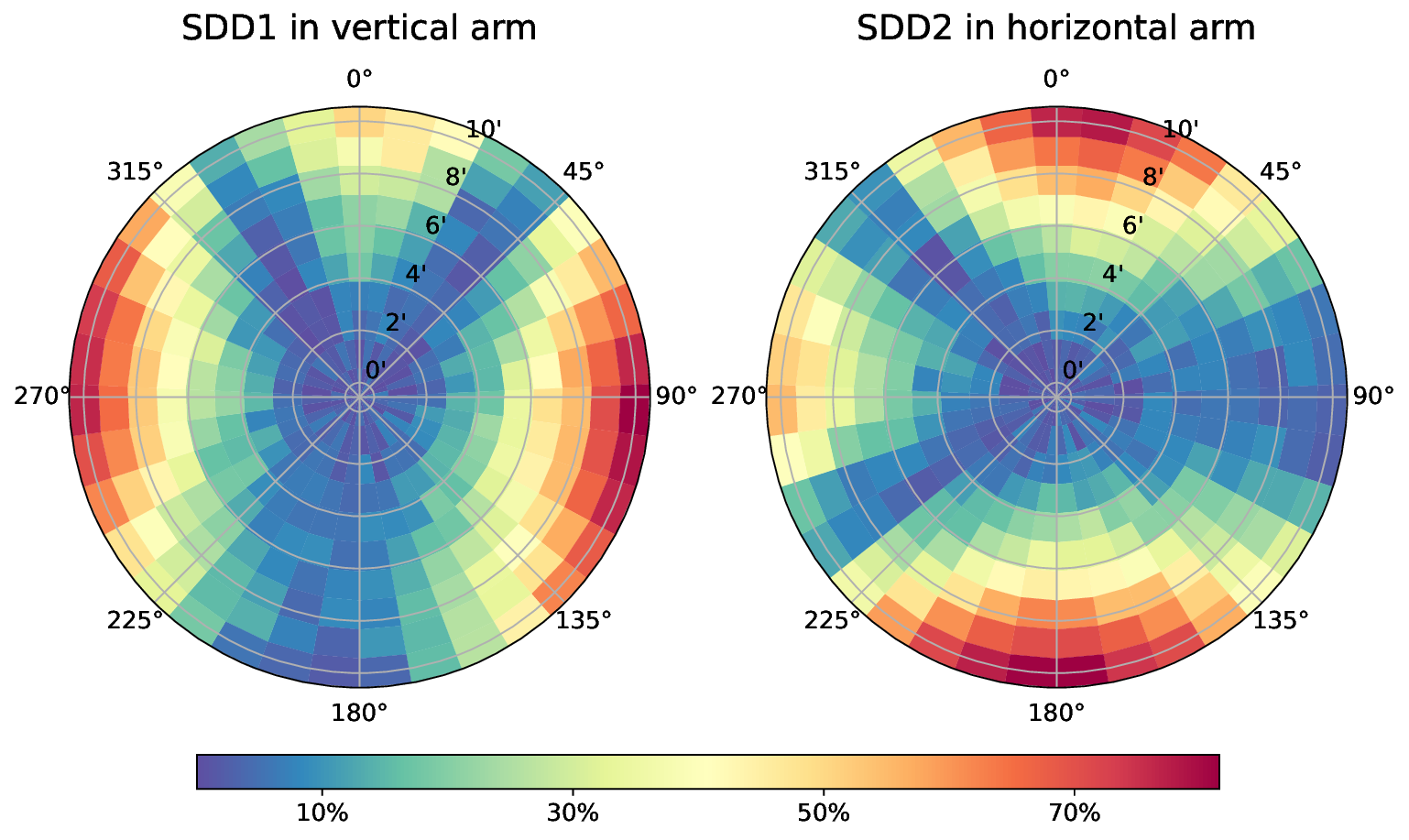}
\caption{Relative-count variations in SDD1 (left panel) and SDD2 (right panel) for the target source at different off-axis angles and directions. }\label{fig_position}
\end{figure}

In the subsequent simulations, we positioned the target source at various off-axis angles and directions, with the off-axis angle ranging from 0 to $\ang{;10;}$, and the off-axis direction ranging from \ang{0;;} to \ang{360;;}. The changes of the relative counts in the two cross-arms detectors under various off-axis angles and off-axis directions are shown in Figure \ref{fig_position}, with brighter color indicating a larger change. It can be observed that for SDD1, the significant change in the relative count occurs when the off-axis direction of the target source is around \ang{90;;} and \ang{270;;}. This is attributed to a portion of the vertical arm moving out of SDD1 at these angles, resulting in a significant decrease in the count and a substantial change in the relative count. Additionally, when the off-axis direction is around \ang{0;;} and the off-axis angle exceeds $\ang{;8;}$, the relative count of SDD1 also undergoes a noticeable change. This is because the portion of the cross-arms closer to the central focal spot has larger counts. As the focal spot shifts towards SDD1 in this off-axis direction, the count in SDD1 increases substantially. Similarly, for SDD2, the change of the relative counts is most pronounced when the target source is off-axis in the vertical direction, i.e. \ang{0;;} or \ang{180;;}, and when it is off-axis in the direction where the horizontal arm detector is located, i.e. \ang{270;;}.

\begin{figure}[ht]%
\centering
\includegraphics[width=0.45\textwidth]{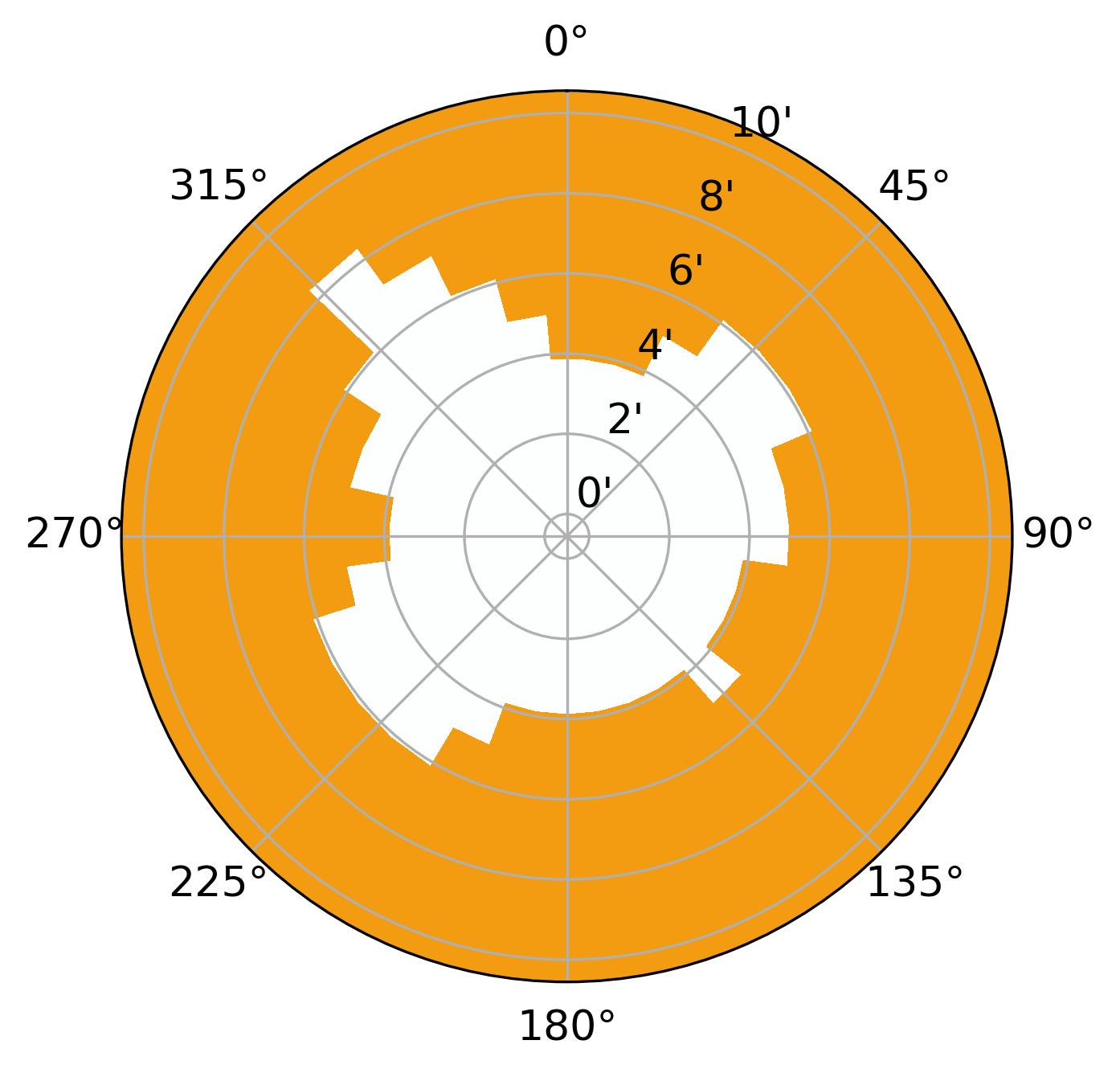}
\caption{Grids colored in orange indicate that target sources from these off-axis angles and off-axis directions can cause the change of the relative count in either of the two cross-arms detectors to exceed 5$\sigma$.}\label{fig_position5sigma}
\end{figure}

We plotted the cases where the changes in two cross-arms detectors exceeded 5$\sigma$ in the polar coordinate system, as shown in Figure \ref{fig_position5sigma}. It can be seen that when observing a Crab-like source with the flux of 1\,Crab, the use of two cross-arms detectors to assist in positioning can improve the positioning accuracy of the Pathfinder of CATCH Type-A to $\ang{;6;}$. The ability to locate sources near the direction of $\ang{45;;}$, $\ang{225;;}$, or $\ang{315;;}$ is weaker compared to $\ang{135;;}$. This is because significant changes in these directions arise from a portion of the vertical or horizontal arm moving out of SDD1 or SDD2; when the off-axis direction of the target source is $\ang{135;;}$, the portion of the cross-arms detected by SDD1 and SDD2 are both farther from the focal spot compared to those detected at on-axis incidence, leading to a greater reduction in the count in SDD1 or SDD2 and a more pronounced change. When the off-axis angle of the target source exceeds $\ang{;10;}$, a portion of the focal spot shifts out of SDD0, and the increase in relative counts on the cross-arms detectors is used to infer the position of the target source. When the focal spot entirely shifts away from SDD0, the localizable range of directions is restricted to ones that are nearly horizontal or vertical.

\begin{figure}[htbp]
\centering
    \begin{minipage}{0.315\textwidth}
        \includegraphics[width=\textwidth]{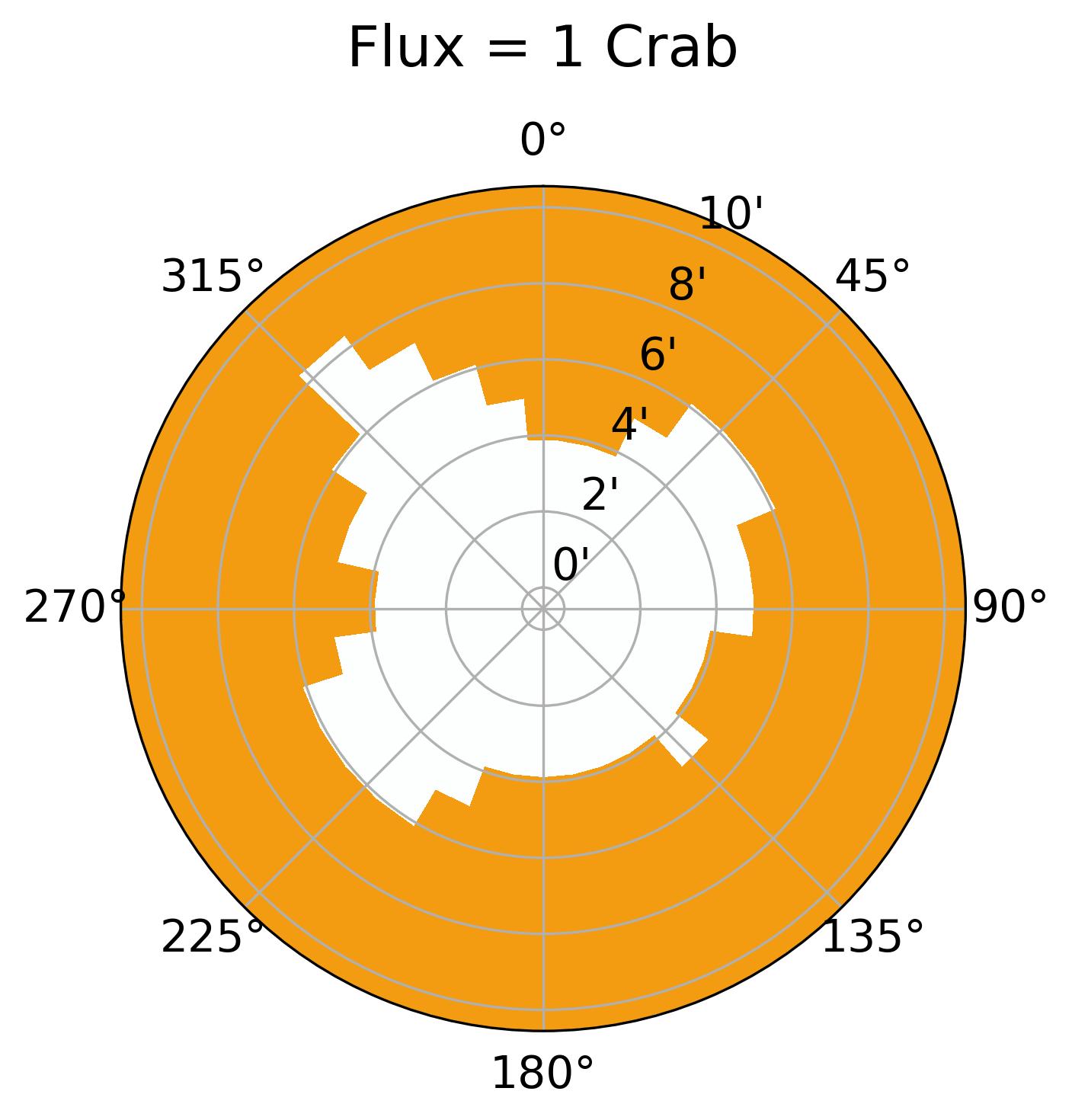}
    \end{minipage}
    \hspace{0.05\textwidth}
    \begin{minipage}{0.315\textwidth}
        \includegraphics[width=\textwidth]{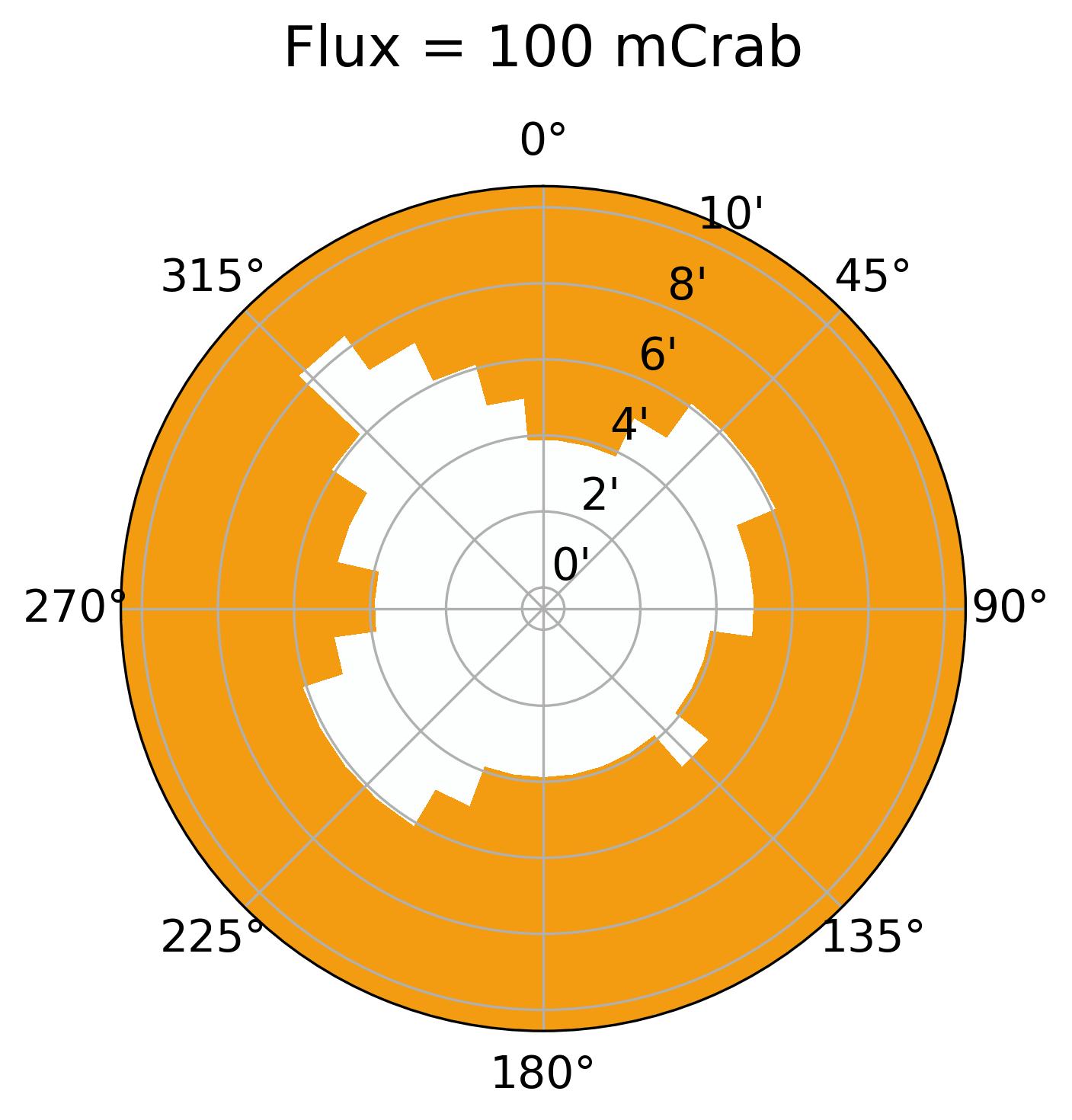}
    \end{minipage}
    \vspace{2em} 
    \begin{minipage}{0.315\textwidth}
        \includegraphics[width=\textwidth]{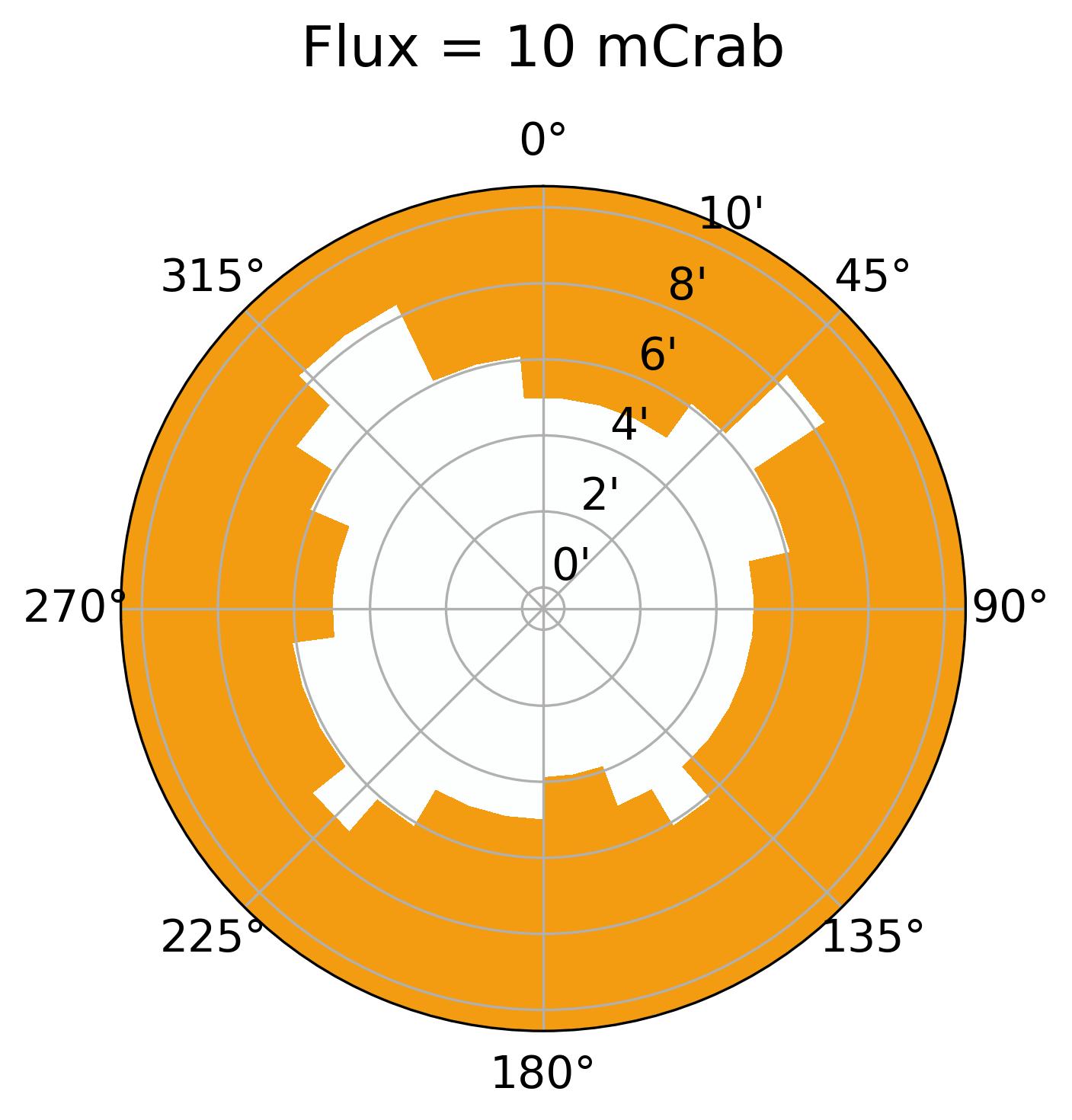}
    \end{minipage}
    \hspace{0.05\textwidth}
    \begin{minipage}{0.315\textwidth}
        \includegraphics[width=\textwidth]{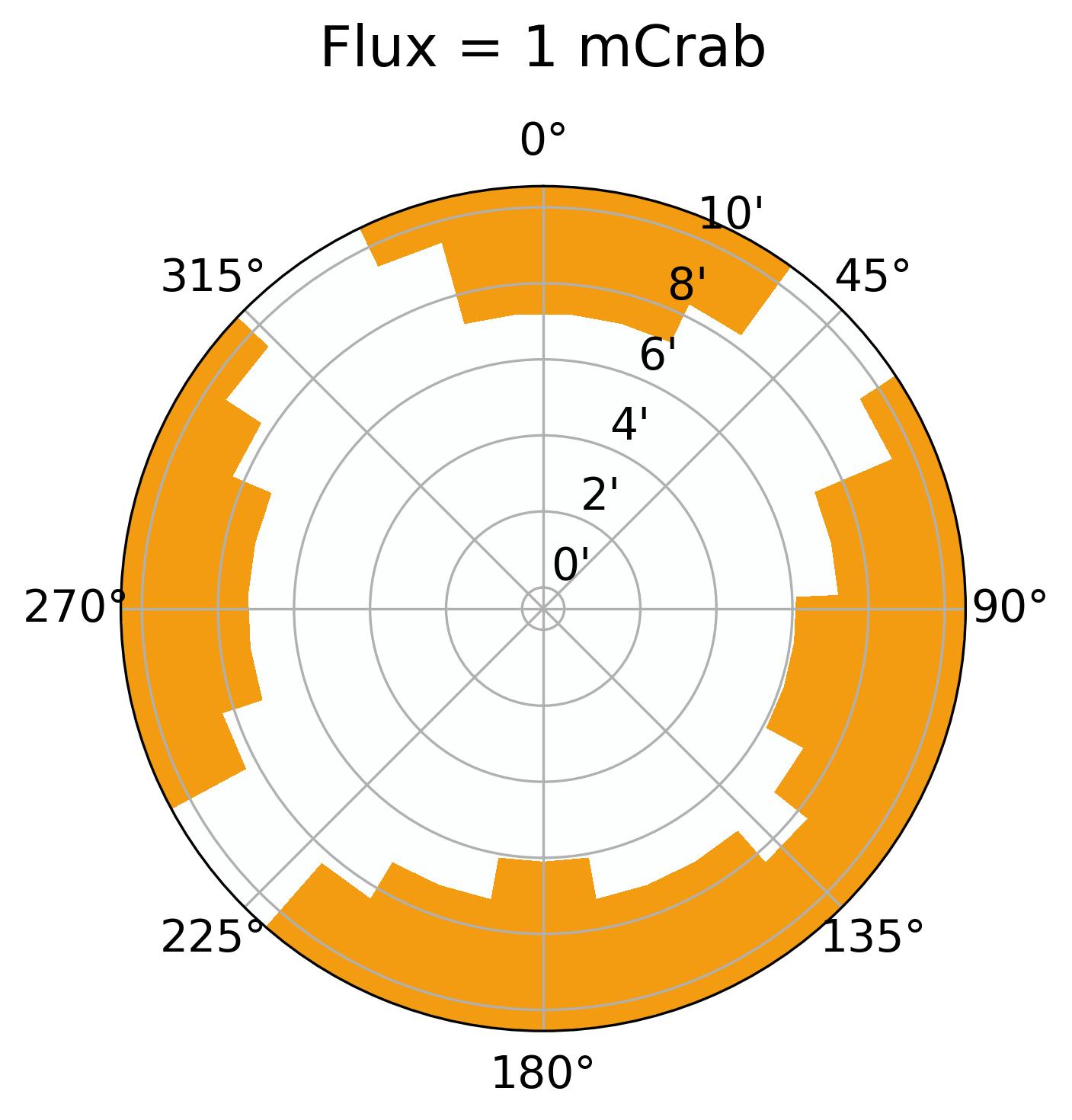}
    \end{minipage}
    \caption{Cases in which the variation in either of the two cross-arm detectors exceeds 5$\sigma$, shown for four flux levels (indicated above each panel). Background counts are retained for the top-row cases and subtracted for those in the bottom row.}
    \label{fig_positionflux}
\end{figure}

Figure \ref{fig_positionflux} shows the detector system’s ability to localize target sources at four distinct flux levels. Localization performance scales with flux in the same manner as contamination identification performance, and, as discussed in Section~\ref{subsec3}, this behavior holds for sources of varying spectra.

In the final design of the Pathfinder of CATCH Type-A, we adopted the updated layout for the detector system.

\section{Powerful SDD array}\label{sec6}

\begin{figure}[ht]%
\centering
\includegraphics[width=0.55\textwidth]{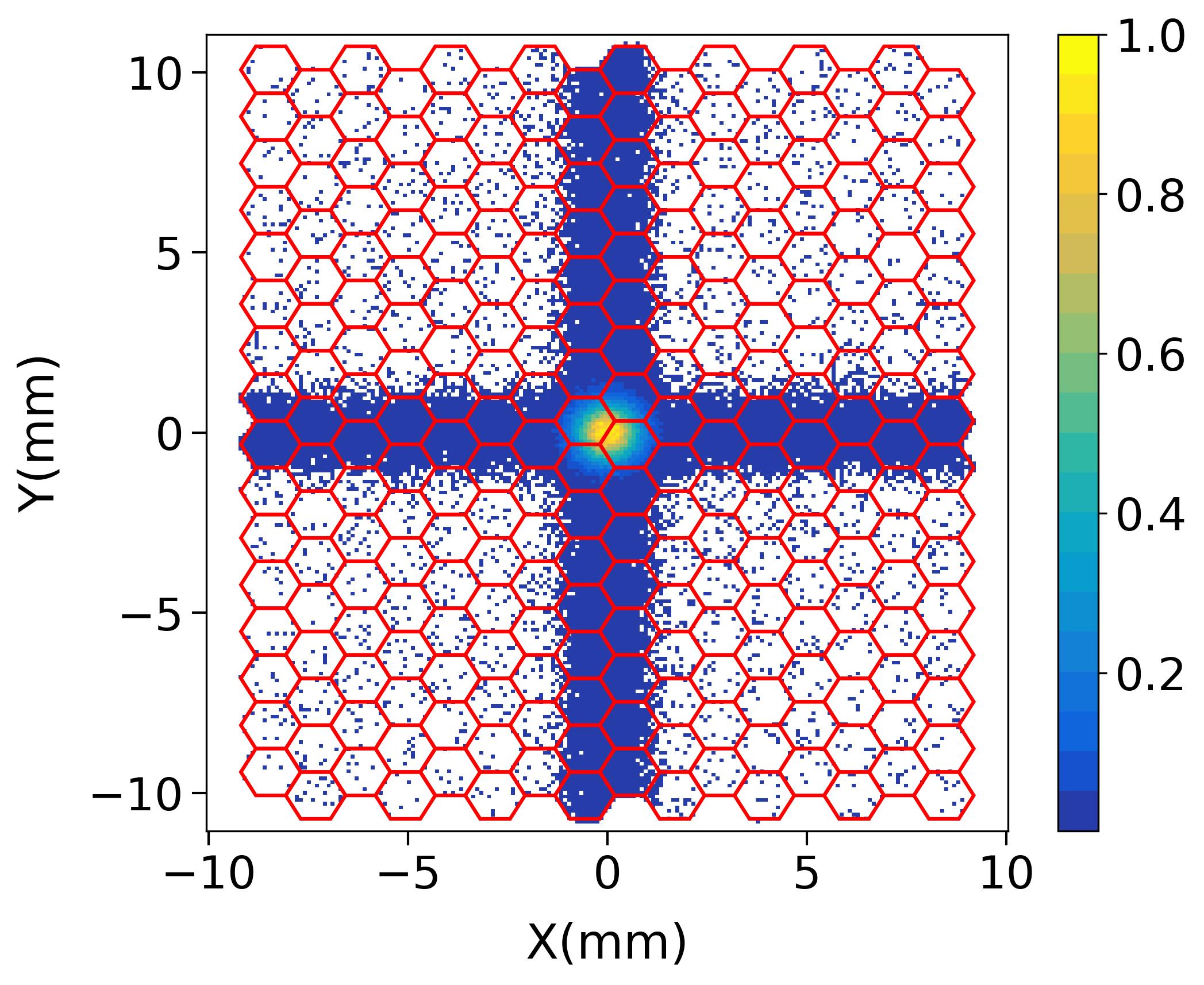}
\caption{Layout of the 16$\times$16 SDD array and the count distribution in it.}\label{fig15}
\end{figure}

In updated plans for CATCH Type-A satellites, the SDD array is expected to be used to achieve a considerable improvement in the capability of identifying contaminating sources and locating target sources. We designed a 16$\times$16 SDD array and integrated it into the Geant4 satellite model that retains the 4$\times$4 MPO configuration of the CATCH Type-A Pathfinder. The SDD array consists of 256 ortho-hexagonal SDDs with a side length of 0.75\,mm, providing a total geometry area is 3.74\,cm$^2$. The SDD array is implemented on a planar surface. Its geometric center coincides with the ideal focal point on the spherical focal surface, which has a radius of curvature of 1\,m. The maximum in-plane distance between the outermost SDD and this center is 13.27\,mm, which corresponds to a maximum deviation of 0.09\,mm from the ideal spherical focal surface. The magnitude of this deviation has a negligible impact on the focusing performance. The simulated count distribution in the SDD array is presented in Figure \ref{fig15}.

\begin{figure}[ht]%
\centering
\includegraphics[width=1.0\textwidth]{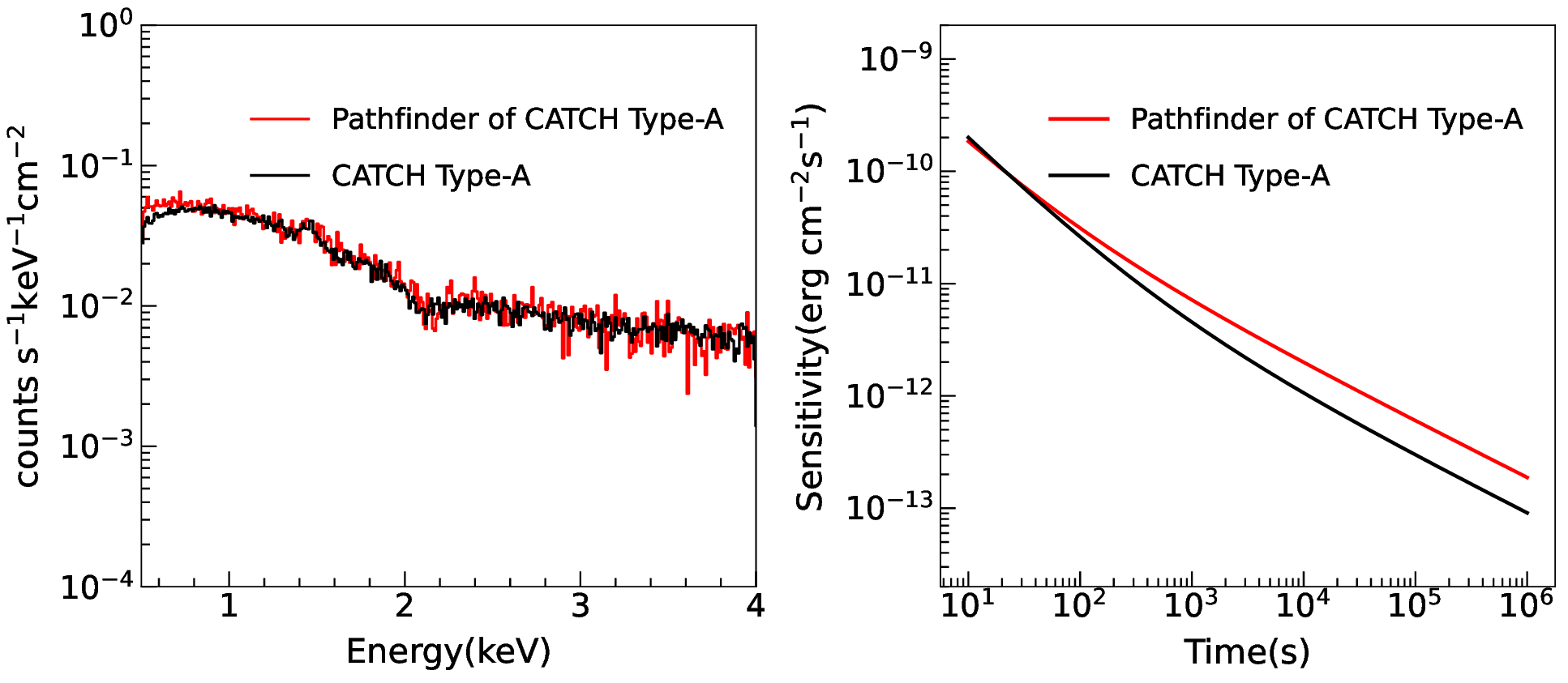}
\caption{Left panel: background spectra for CATCH Type-A and its Pathfinder. Right panel: 0.5--4\,keV sensitivity curves for both configurations.}\label{fig16}
\end{figure}

We conducted a simulation study on the background in the CATCH Type-A payload configuration and evaluated its sensitivity, following the same approach as in \cite{huang} for the CATCH Type-A Pathfinder. The top panel of Figure \ref{fig16} shows the total background energy spectrum of CATCH Type-A and compares it with that of the CATCH Type-A Pathfinder. The two configurations exhibit similar background spectra. The minor difference at the lowest energy end arises from differences in the detector's sensitive layer: the Pathfinder features a 450\,$\upmu$m-thick Si layer without any additional materials above it, while CATCH Type-A has a 300\,$\upmu$m-thick Si layer covered by 100\,nm of Al and 50\,nm of SiO$_2$. Consequently, low-energy particles are partially absorbed by the Al and SiO$_2$ layers before reaching the sensitive layer of CATCH Type-A. The total background of CATCH Type-A is $2.48 \times 10^{-1}$\,counts~s$^{-1}$, i.e., $6.63 \times 10^{-2}$\,counts~s$^{-1}$~cm$^{-2}$ after area normalization. The bottom panel of Figure \ref{fig16} plots the sensitivity of CATCH Type-A in the energy range of 0.5--4\,keV, also in comparison to the Pathfinder. With a 1000\,s exposure time, the sensitivity of CATCH Type-A can reach $4.5 \times 10^{-12}$\,erg~cm$^{-2}$~s$^{-1}$, which is better than that of the Pathfinder. In calculating sensitivity, we focused on the detectors covering the focal spot. For the CATCH Type-A Pathfinder, this corresponds to the central detector, while for CATCH Type-A, it involves the four central pixels. The geometry areas, background counts, and source counts detected per unit time for these two detector regions are listed in Table \ref{tab_sensitivity}. The geometry area of the four central pixels on CATCH Type-A is significantly reduced compared to that of the primary detector on the CATCH Type-A Pathfinder. Since the background is distributed over a large area, the background count for CATCH Type-A is also considerably decreased. The source count is concentrated at the focal spot, resulting in a similar source count for both configurations. Therefore, CATCH Type-A demonstrates better sensitivity.

\begin{table}[h]
\caption{Geometric areas, background count rate, and source count rate for the Pathfinder’s primary detector and CATCH Type-A’s four central pixels.}
\label{tab_sensitivity}
\begin{tabular}{@{}ccccccc@{}}
\toprule
       &  Area & Background count rate & Source count rate \\
       &  (mm$^2$) & (cts~s$^{-1}$)  & (cts~s$^{-1}$) \\       
\midrule
Pathfinder of CATCH Type-A &  50  &  0.049  &  208    \\
CATCH Type-A & 5.85 &  0.006   & 152 \\
\botrule
\end{tabular}
\end{table}

\begin{figure}[ht]%
\centering
\includegraphics[width=0.89\textwidth]{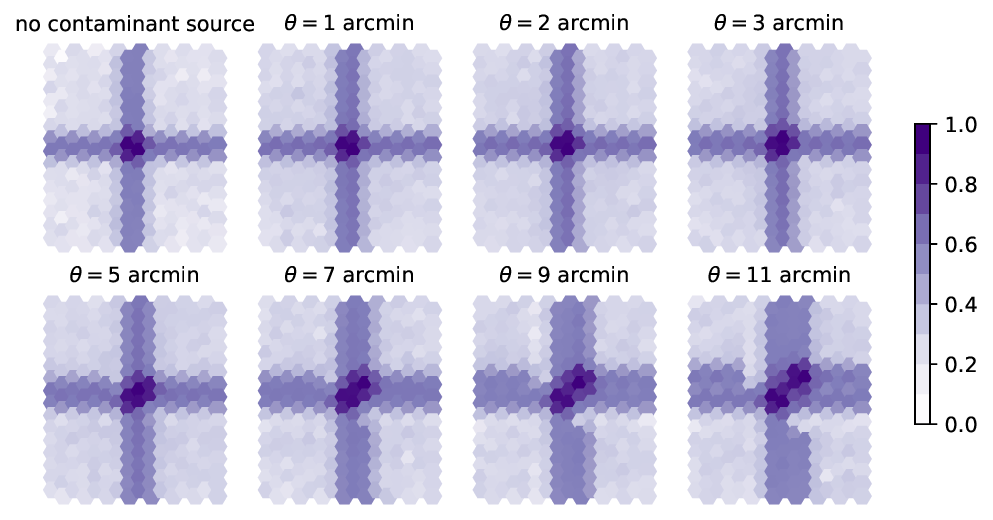}
\caption{Top-left: count distribution in the SDD array without contaminating source. Remaining panels: count distributions with contaminating source at increasing off-axis angles (labeled above each image). Color intensity scales with the logarithm of the count in each SDD pixel.}\label{fig_SDDarray_contamination}
\end{figure}

To evaluate CATCH Type-A's ability to identify contaminating sources, we simulated and compared the count distribution in the SDD array with and without contaminating sources present during observation. Figure~\ref{fig_SDDarray_contamination} shows several cases where the target source is on-axis, while the contaminating source is incident from an off-axis direction of \ang{45;;} but with varying off-axis angles. In this figure, the color intensity represents the count in each SDD pixel. Given the large disparity in counts between the focal spot and off-spot pixels, the color intensity is proportional to the logarithm of the count to enhance visibility of the cross-arms. The logarithmic counts are normalized when mapped to color. As seen in Figure~\ref{fig_SDDarray_contamination}, the two focal spots become increasingly separated as the off-axis angle increases.

\begin{figure}[h]
\centering
\begin{minipage}{0.48\textwidth}
\centering
\includegraphics[width=\textwidth]{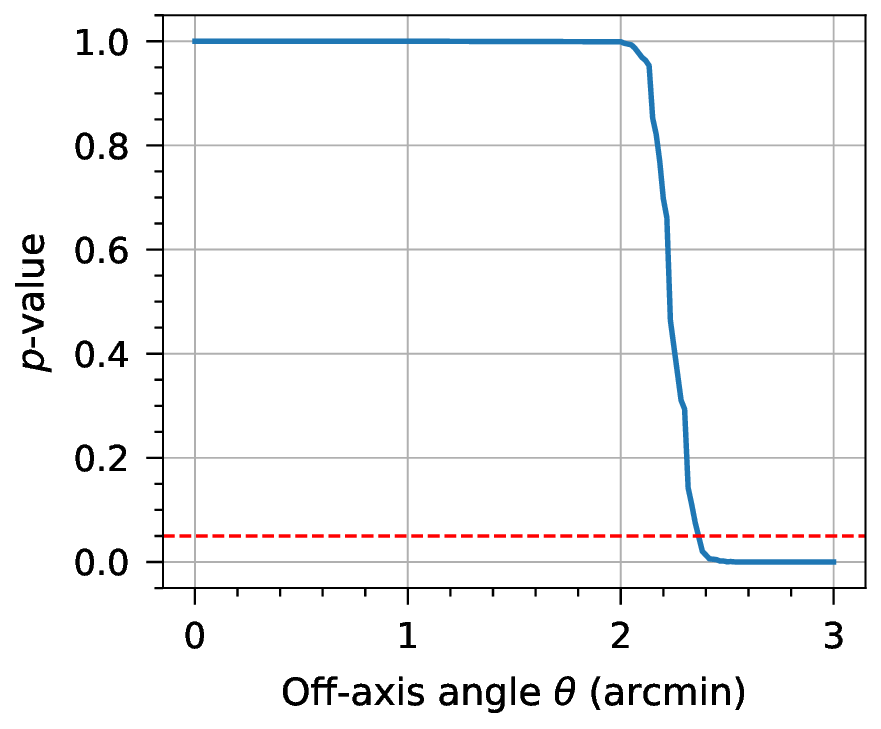}
\end{minipage}
\begin{minipage}{0.48\textwidth}
\centering
\includegraphics[width=\textwidth]{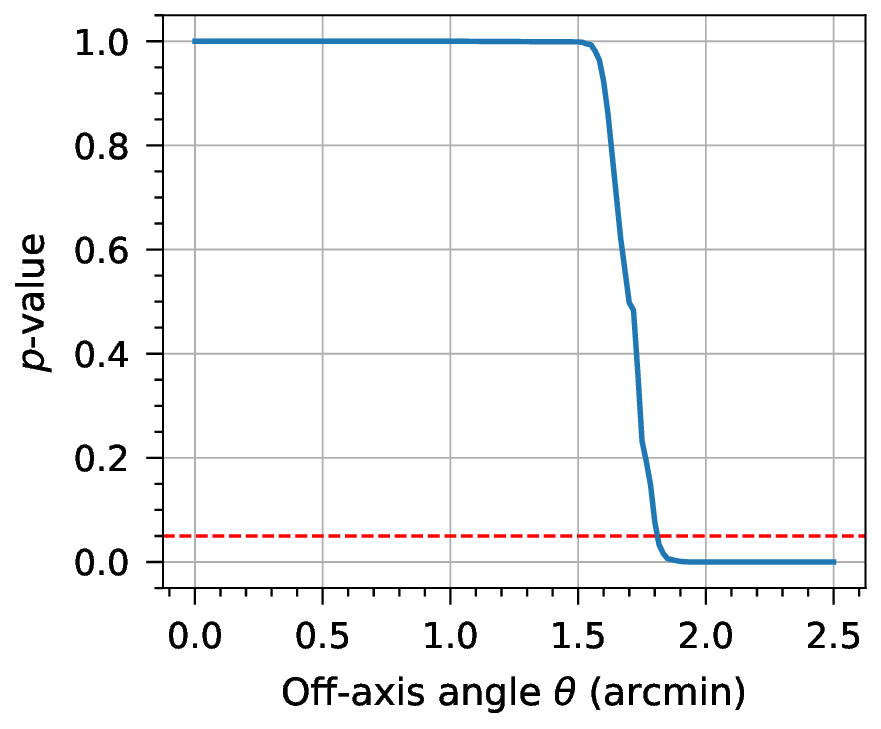}
\end{minipage}
\caption{Left panel: \(p\)-value variation as a function of the contaminating source's off-axis angle for a 1\,Crab target and a 1\,Crab contaminating source with 1\,s exposure. Right panel: \(p\)-value variation as a function of the off-axis angle of the target source for a 1\,Crab source with 1\,s exposure. The dashed red line marks the \(p=0.05\) significance threshold.}
\label{fig_pvalue_distribution}
\end{figure}

To quantify the difference of count distribution and identify contaminating sources, we employed the chi-square goodness-of-fit test. Specifically, we treated the count distribution in the SDD array without a contaminating source as the ideal distribution \(E\) and that with a contaminating source as the observed distribution \(O\). For each case, the chi-square value was computed as:
\begin{equation}
\chi^{2}_{\rm obs} = \sum_{i=1}^{256} \frac{(O_i - E_i)^2}{E_i},
\end{equation}
where \(E_i\) and \(O_i\) represent the counts in the \(i\)-th SDD pixel in ideal and observed cases, respectively. The corresponding \(p\)-value was obtained from the chi-square survival function \(p=P(\chi^2_\nu \geq \chi^2_{\rm obs})\), with \(\nu=255\) degrees of freedom. These calculations were performed using the chisquare function from the SciPy statistics package in Python \cite{2020SciPy-NMeth}. A smaller \(p\)-value indicates that the two distributions are statistically distinguishable. The left panel of Figure~\ref{fig_pvalue_distribution} shows how the \(p\)-value varies with the off-axis angle of the contaminating source, assuming both the target and contaminating sources have a flux of 1\,Crab and an exposure time of 1\,s. The \(p\)-value remains close to unity for small angles, then decreases sharply beyond about \ang{;2;}, and falls below the significance threshold of \(p=0.05\) near \ang{;2.4;}. This indicates that CATCH Type-A can effectively identify a contaminating source of comparable brightness when its off-angle angle greater than \ang{;2.4;}.

\begin{figure}[ht]%
\centering
\includegraphics[width=0.89\textwidth]{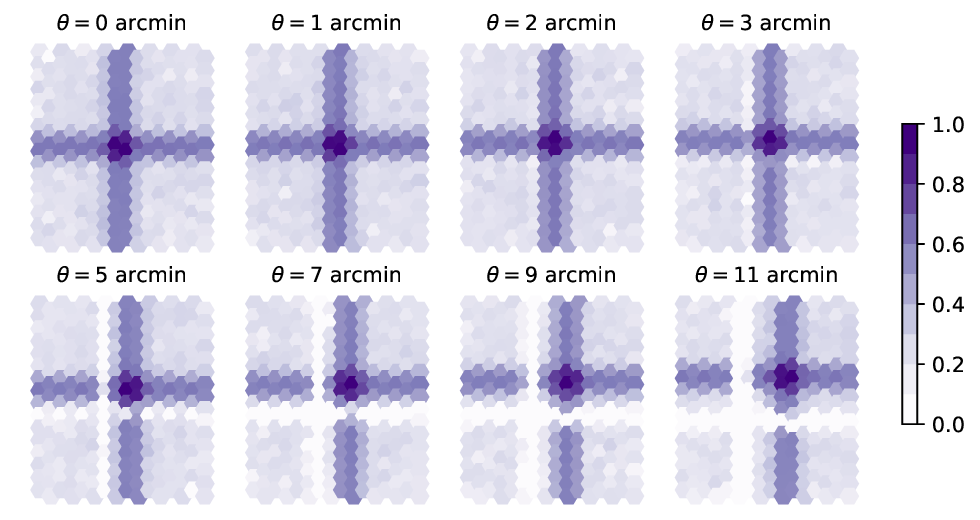}
\caption{Top-left: count distribution for on-axis incidence. Remaining panels: Count distributions for off-axis incidence with increasing off-axis angles (labeled above each image). Color intensity scales with the logarithm of the count in each SDD pixel. White pixels indicate shadows cast by the optical system’s supporting frame.}\label{fig_SDDarray_position}
\end{figure}

We also assessed CATCH Type-A's positioning accuracy via simulation. Figure \ref{fig_SDDarray_position} shows the simulated count distribution in the SDD array for on-axis and several off-axis target-source incidences. In the off-axis cases, the off-axis direction is fixed at \(\ang{45}\), but the off-axis angles vary. As the off-axis angle increases, the focal spot and cross-arms in the SDD array visibly shift toward the upper right. We used the chi-square goodness-of-fit test to quantify the differences in the SDD array's count distribution between off-axis and on-axis target-source incidences. The resulting \(p\)-values as a function of off-axis angle is plotted in the right panel of Figure~\ref{fig_pvalue_distribution}. For a 1\,Crab with 1\,s exposure time, the \(p\)-value drops below 0.05 at an off-axis angle of about \ang{;1.8;}, indicating that CATCH Type-A can achieve a positioning accuracy of \ang{;1.8;} under these conditions.

Here, we demonstrate the contaminating source identification and source localization capabilities achieved by combining a 16\,$\times$\,16 SDD array with a 4\,$\times$\,4 MPO pieces. It is worth noting that the use of a 16\,$\times$\,16 SDD array has significantly enhanced these capabilities compared to the original CATCH Type-A satellites described in \cite{licatch}, as well as the Type-A Pathfinder. This highlights the attractive scientific prospects of CATCH as an astronomical constellation mission that continues to upgrade in response to scientific requirements and technological developments.


\section{Conclusion}\label{sec7}

The CATCH Type-A Pathfinder is designed to conduct on-orbit technology validation for the timing satellites (Type-A) within the CATCH constellation and to perform timing observations of transients. It is equipped with MPOs featuring a cross-shaped PSF and 4-pixel SDDs. In this work, we conducted simulated observations in Geant4 to investigate the capability of the CATCH Type-A Pathfinder in identifying contaminating sources and locating target sources. 

Through simulated observations of several scenarios where two sources are present within the FOV—one as the target source and the other as the contaminating source—we proposed a detector layout that can effectively enhance the capability to identify contaminating sources. This layout places one detector at the focal spot as the primary detector, one detector at the vertical arm, one detector at the horizontal arm, and the remaining detector at the diffuse patch. We simulated cases where contaminating sources originated from various off-axis angles and directions and compared the relative counts in the cross-arm detectors with those in the absence of the contaminating source. The results indicate that this detector layout, when used in conjunction with MPOs, allows for the identification of contaminating sources with an off-axis angle larger than $\ang{;8;}$ if the flux for both the target source and contaminating source is 1 Crab and the exposure time is 200\,s.

The cross-arm detectors can also assist in source positioning. We simulated observations of target sources from various off-axis angles and directions and compared the relative counts in the cross-arm detectors with those for on-axis incidence. The results demonstrate that for a target source with a flux of 1\,Crab and an exposure time of 200\,s, the CATCH Type-A Pathfinder can achieve a positioning accuracy of $\ang{;6;}$.

The CATCH constellation plans to integrate MPOs with a multi-pixel SDD array in CATCH Type-A satellites. We evaluated the background and sensitivity of the payload configuration combining a 16\,$\times$\,16 SDD array with a 4\,$\times$\,4 MPO array and compared them with the CATCH Type-A Pathfinder. Their background spectra per unit area differ slightly at the lowest energy end. The sensitivity of the upgraded configuration is better than that of the CATCH Type-A Pathfinder. Equipped with the SDD array, for an exposure time of 1\,s, the contaminating sources can be identified with an off-axis angle exceeding \ang{;2.4;} when the flux of both the target and contaminating source is 1\,Crab, and the positioning accuracy of a source with a flux of 1 Crab can reach $\ang{;1.8;}$. Multi-pixel SSD arrays are now in active development, and a key attraction of the CATCH mission is its continual upgrade according to new detection technologies.

This study establishes a foundational framework for identifying the presence of contaminating sources, which serves as the basis for our future data processing and analysis. After launch, the telescope will be calibrated in orbit, and the measured count distribution can be used to update the model and further validate the method. We will pursue more in-depth research, with plans to develop observation strategies for practical observations to achieve a more comprehensive and precise identification of these sources.


\section*{Acknowledgments}
We would like to thank all the members of the CATCH team. This work is supported by the National Natural Science Foundation of China under the Grant Nos. 12122306, 12003037, and 12173056.

\input{sn-article.bbl}

\end{document}